\documentclass[twocolumn,prb,aps,nobalancelastpage,amsfonts]{revtex4}

\usepackage{graphicx}
\DeclareGraphicsExtensions{.eps}
\begin{document}

\title{Damping of the de Haas-van Alphen oscillations in the superconducting state of MgB$_2$}
\author{J.D. Fletcher and A. Carrington}
\affiliation{H.H. Wills Physics Laboratory, University of Bristol, Tyndall Avenue, BS8 1TL, United Kingdom.}

\author{S.M. Kazakov and J. Karpinski}
\affiliation{Laboratorium f\"{u}r Festk\"{o}rperphysik, ETH Z\"{u}rich, CH-8093 Z\"{u}rich, Switzerland.}
\date{\today}

\begin{abstract}
The de Haas-van Alphen (dHvA) signal arising from orbits on the $\pi$ Fermi surface sheet of the two-gap superconductor
MgB$_2$ has been observed in the vortex state below $H_{c2}$. An extra attenuation of the dHvA signal, beyond those
effects described in the conventional Lifshitz-Kosevich expression, is seen due to the opening of the superconducting
gap. Our data show that the $\pi$ band gap is still present up to $H_{c2}$. The data are compared to current theories
of dHvA oscillations in the superconducting state which allow us to extract estimates for the evolution of the $\pi$
band gap with magnetic field.  Contrary to results for other materials, we find that the most recent theories
dramatically underestimate the damping in MgB$_2$.
\end{abstract}

\pacs{}%
\maketitle

\section{Introduction}
The existence of two distinct energy gaps in the superconducting state of MgB$_2$
has been demonstrated by a number of experiments, including:
tunneling,\cite{IavaroneKKKCHKCKKL02} specific
heat,\cite{BouquetWFHJJP01,BouquetWSPJLT02} magnetic penetration depth
\cite{ManzanoCHLYT02} and angle resolved photoemission spectroscopy (ARPES).
\cite{SoumaMSTMWDKCSK03,TsudaYTKMYIHS03} Although there have been indications of
two, or multi-gap, effects in other materials, \cite{ShenSP65,BinnigBHB80} it is in
MgB$_2$ where the behavior has been most thoroughly investigated. Band structure
calculations \cite{KortusMBAB01} have shown that the Fermi surface of MgB$_2$
consists of four sheets, mostly arising from the boron orbitals. Two
quasi-two-dimensional sheets originate from the boron $\sigma$ orbitals and two more
isotropic honeycomb shaped sheets from the boron $\pi$ orbitals. Theoretical work
\cite{LiuMK01,ChoiRSCL02} has predicted that the superconducting energy gap is
substantially larger on the $\sigma$ sheets than on the $\pi$ sheets (at zero field
and temperature the two gaps have been measured to be $\Delta_\sigma\simeq$ 78~K and
$\Delta_\pi\simeq$ 29~K).\cite{ManzanoCHLYT02,CarringtonM03}

Although there have been several studies which have accurately measured the
temperature dependence of the two
gaps,\cite{GonnelliDUSJKK02,IavaroneKKKCHKCKKL02,SzaboSKKMFMMJ01} there has been
much less work in establishing how they evolve with magnetic field. Theoretical work
\cite{GraserDS04} predicts that at high field both gaps decrease towards zero at a
common upper critical field ($H_{c2}$) value, although at lower field $\Delta_\pi$
is depressed much more rapidly than $\Delta_\sigma$, particularly for $H\|ab$.
\cite{dahmprivate} Experimentally, the only direct studies of this have been by
point contact spectroscopy. It has been suggested \cite{DagheroGUSJKK03} that
$\Delta_\pi$ goes to zero at $\sim 1~\rm{T}$ (for $H\|c$), which is much lower than
$H_{c2}$, although others\cite{BugoslavskyMPCCPX03,Gonnelli03} have concluded that
$\Delta_\pi$ remains finite above $1~\rm{T}$ but becomes increasingly difficult to
resolve because of scattering.  In this paper, we use dHvA measurements as a probe
of the gap in high magnetic fields. Our data clearly show the presence of a gap on
one $\pi$-band sheet right up to the `bulk' $H_{c2}$.

The existence of dHvA oscillations in the superconducting state has been the subject of study for a number of years,
and has been observed in a number of different materials: NbSe$_2$ (Ref. \onlinecite{CorcoranMOPSTHGG94}), V$_3$Si
(Ref. \onlinecite{CorcoranHHMSV94}), Nb$_3$Sn (Ref. \onlinecite{HarrisonHMSVM94}), $\kappa-$(ET)$_2$Cu(NCS)$_2$ (Ref.
\onlinecite{ClaytonIHMSS02}), YNi$_2$B$_2$C (Ref. \onlinecite{HeineckeW95}) and several heavy fermion
compounds.\cite{SettaiSIMNAHHO01,InadaYHSTHYOY99,OhkuniITSSHHYOYTY99,BergemannJMHLLBF97,HedoIIYHOHH95} The effect is
usually thought to arise through the overlap of quasiparticle states outside the vortex core. The amplitude of the
oscillations is governed by the magnitude of the field dependent superconducting energy gap. In theory the effect can
be used to resolve the gaps on different Fermi surface sheets and to probe the field dependence of these gaps. However,
in the current study scattering restricts our study to only one Fermi surface sheet.

The amplitude of the dHvA oscillations is interpreted using the Lifshitz-Kosevich equation for the oscillatory torque
$\Gamma$ of a three-dimensional Fermi liquid. The amplitude of the first harmonic is given by
\cite{shoenberg,WassermanS94}
\begin{equation}
\Gamma_{osc}\propto
\frac{B^\frac{3}{2}}{[\mathbb{A}^{\prime\prime}]^\frac{1}{2}}\frac{dF}{d\theta}R_DR_TR_SR_{SC}\sin\left[\frac{2\pi
F}{B}+\varphi\right].\label{lkeq}
\end{equation}
Here the dHvA frequency $F$ is related to the extremal area ($\mathbb{A}$) of the orbit in $k$-space by $F=(\hbar /2\pi
e)\mathbb{A}$, $\mathbb{A}^{\prime\prime}=\partial^2\mathbb{A}/\partial k^2$ is the curvature factor and $\varphi$ is
the phase. The factor $R_T = X/(\sinh X)$ where $X = \frac{2\pi^2k_{_B}}{\hbar e} \frac{m^*T}{B}$, accounts for the
effects of thermal broadening of the Landau-levels. It is from this temperature dependent term that the quasi-particle
effective mass $m^*$ is determined. The Dingle factor accounts for the effect of impurities and is given by $R_D =
\exp(-\pi/\omega_c\tau)$, where $\omega_c=eB/m_B$, $m_B$ is the unenhanced band mass \cite{shoenberg,WassermanS94} and
$\tau$ is the scattering time. The spin splitting factor $R_S$ accounts for the reduction of amplitude caused by
beating between the spin-up and spin-down Fermi surfaces. It is given by $R_S = \cos(\pi n g m_B(1+S)/2m_e)$ where $S$
is the orbitally averaged exchange-correlation (Stoner) enhancement factor, $g$ is the electron $g$-factor, $m_{e}$ is
the free-electron mass and $n$ is an integer. The final factor, $R_{SC}$, parameterizes the effect of superconductivity
on the dHvA amplitude. It is this factor which is the primary focus of the current work and will be described more
fully below.

\begin{figure} \includegraphics*[width=6cm]{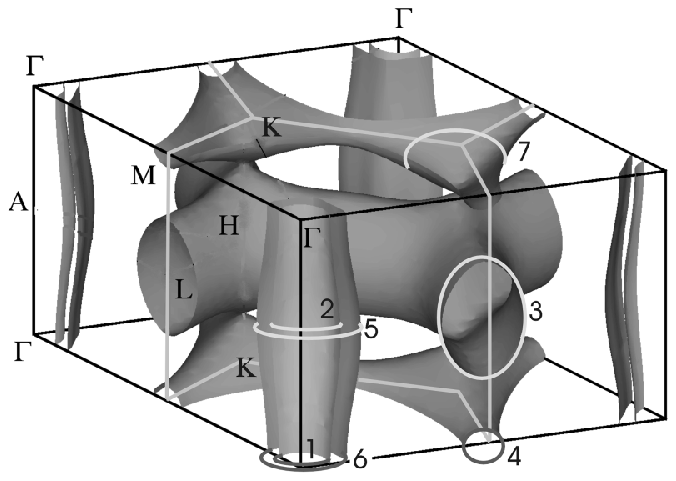} \caption{Fermi surface of MgB$_2$ with predicted dHvA orbits
 (for frequencies less than 10kT).  The orbit which survives into the
superconducting state, and is the subject of the majority of the present study, is orbit 3. [Figure adapted from Kortus
\textit{et al.} \protect\cite{KortusMBAB01}]} \label{fsfig} \end{figure}

The calculated Fermi surface of MgB$_2$, with predicted\cite{Harima02,RosnerAPD02,MazinK02} dHvA orbits, is shown in
Fig. \ref{fsfig}.  We denote the frequency of each orbit, $n$, at a general angle by $F_n$. Previous
studies\cite{YellandCCHMLYT02,CarringtonMCBHYLYTKK03} have succeeded in observing orbits 1 to 6, thus verifying the
topology of the calculated Fermi surface.  The measured $k$-space areas of the orbits and the quasiparticle effective
masses were found to be in good overall agreement with the calculations.\cite{Harima02,RosnerAPD02,MazinK02} Orbit 7
(Ref.\ \onlinecite{Harima02}) was not observed experimentally, perhaps because of a slight departure of Fermi surface
topology from the calculations, or alternatively because of increased scattering on this orbit.

\begin{figure} \includegraphics*[width=8cm]{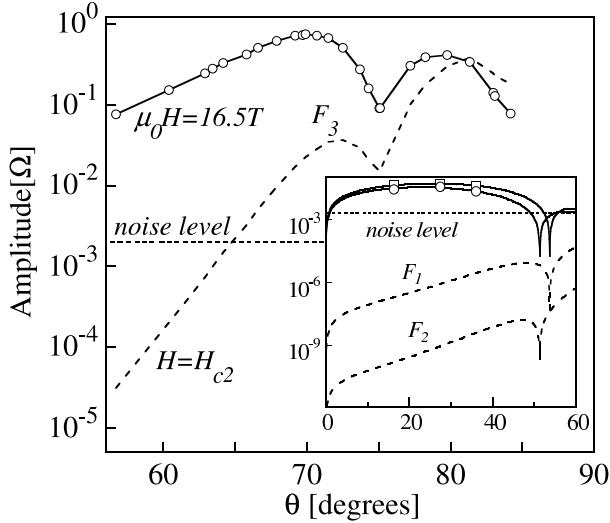} \caption{The main figure shows the amplitude of the oscillations for orbit 3 at 16.5T (symbols) as a
function of angle ($\theta$) (the solid line is a guide to the eye). The dashed line shows the extrapolated amplitude
at $H=H_{c2}(\theta)$ using the measured Dingle factor $R_D$. The inset shows the corresponding angular dependence of
the amplitude of orbits 1 and 2 at 16.5~T fitted according to Eq. \ref{lkeq} (solid lines), and the extrapolation to
$H=H_{c2}(\theta)$ (dashed lines). The noise level of our measurement circuit is indicated on both parts of the figure.
The mean-free-paths were measured to be 500\AA, 610\AA~ and 840\AA~ for $F_1$, $F_2$ and $F_3$ respectively.}
\label{ampthetafig}
\end{figure}

The relative amplitude of the various orbits is somewhat sample dependent.  We have found that the mean-free-paths for
different orbits does not change in a uniform way between samples.\cite{CarringtonMCBHYLYTKK03} Orbits 1-3 have by far
the largest amplitude for fields below 20~T.  The frequencies of these orbits all vary approximately like
$1/\cos\theta$ (or $1/\sin\theta$), although for $F_1$ and $F_2$ there are some departures from this simple behavior
due to the warping of the $\sigma$ band sheet (see Ref.\ \onlinecite{YellandCCHMLYT02} for details).  The amplitude of
orbits 1 -- 3 as a function of angle $\theta$ as the magnetic field is rotated from $B\|c$ ($\theta$=0) to $B\|a$
($\theta$=90$^\circ$) is shown in Fig.\ \ref{ampthetafig}. The strong angular dependence of $H_{c2}$, combined with the
exponential attenuation of the signal by the Dingle factor, limits the range of angles where we are able to observe
oscillations below $H_{c2}$ to 66$^\circ\lesssim \theta \lesssim $81$^\circ$ for the present crystals. The amplitude of
the signal from orbits 1 and 2 at $H_{c2}$ (Fig.\ \ref{ampthetafig} inset) is many orders of magnitude below our noise
level, and so our study is limited to orbit 3 on the electron-like $\pi$ sheet. We estimate that the mean-free-path
$\ell$ would have to be $\sim$ 3 times larger on orbit 1 for us to be able to see these oscillations at $\mu_0H=4$~T
(for these crystals $\mu_0H_{c2}\simeq 4.3$~T at $\theta=30^\circ$, although this does decrease with increasing $\ell$
[Ref.\ \onlinecite{CarringtonMCBHYLYTKK03}]).

\section{Experimental Details}

Rather than measure the oscillations in the magnetic susceptibility, as in a conventional field modulated dHvA
experiment, we have used miniature piezoresistive cantilevers \cite{piezolevers} to measure the oscillations in the
torque. We have found that this latter technique is significantly more sensitive than the former for the small size of
the available high quality crystals of MgB$_2$.

The MgB$_2$ crystals were fixed with epoxy to the end of a boron-doped silicon cantilever about 150$\mu$m long.
Deflections in the cantilever were measured through the strain-induced changes in its electrical resistance, using an
AC bridge technique. The torque values are reported here in units of bridge resistance $R$, i.e, the off-balance
voltage divided by the excitation current (the change in cantilever resistance is $4R$).  We estimate, using the weight
of the sample, that $\Gamma\simeq10^{-11}$R ($\Gamma$ in Nm and $R$ in $\Omega$). The noise limit is around 2m$\Omega$
or $\sim 10^{-14}$ Nm (this can vary by up to a factor of 10 between different levers). In addition to the changes in
resistance caused by the torque there is a small monotonic magnetoresistance of the sensor which we have not attempted
to correct for.

The cantilever is mounted on a single axis rotation stage in a $^3$He cryostat inside the bore of a 19~T
superconducting magnet (20.5~T at $T$=2.2~K). Unless otherwise stated, all measurements in this paper were performed in
liquid $^3$He at 320$\pm$20~mK. The orientation of the cantilever mount is detected using a pickup coil and a small
modulated field collinear with the DC field.  The small offset between the crystal plane and the cantilever mount was
corrected using the symmetry of the dHvA frequencies around the $\theta=0^\circ$ and 90$^\circ$ points giving an
uncertainty in the out of plane angle of $\pm$0.2$^\circ$ (at $\theta \simeq 90^\circ$). The in-plane orientation of
the crystal, $\phi$, (measured from the \textit{a}-axis) is fixed for each run and measured by optical photographs and
Laue X-ray diffraction (to $\pm 5^\circ$). A more precise determination of $\phi$ was made by comparing the minimum in
the frequency of orbit 3 as a function of $\theta$ to the known frequency for this orbit at $\theta=\phi=0^\circ$, as
measured on a large number of other crystals \cite{CooperCMYHBTLKK03} (the minimum frequency varies approximately as
$1/\cos \phi )$.

The torque, both due to the oscillations and the background magnetism, causes a deflection of the lever, and so
$\theta$ is not quite constant during the field sweep. Using the strong angular dependence of the dHvA frequencies, we
are able to measure the correspondence between the measured torque signal and the angular deflection of the cantilever.
We find that the deflection is given by $\Delta\theta=(0.05\pm0.01)^\circ/\Omega$, which is approximately 4 times
smaller than the value estimated from the manufacturer's data sheet.\cite{piezolevers} This `torque interaction' effect
gives rises to spurious harmonics and mixing of the main frequencies.  As the amplitude of the signal is also angle
dependent (see Fig.\ \ref{ampthetafig}) the field dependence of the signal will not exactly follow Eq.\ \ref{lkeq}.
Fortunately, this effect is small enough to be neglected for the purposes of the current work.  This was verified by
measuring samples with different masses and comparing up and down field sweeps (see later).

The samples used in this work were grown by a high pressure synthesis route as described in Ref.\
\onlinecite{KarpinskiAJKPWRKPDEBVM03}. Most of the work reported here was conducted on a crystal measuring $300 \times
160 \times 30 \mu$m$^3$ (mass=3.9$\mu$g). The $T_c$ of samples from this batch (AN77) was measured to be 38.5~K.

\section{Results}

\begin{figure}
\includegraphics*[width=8cm]{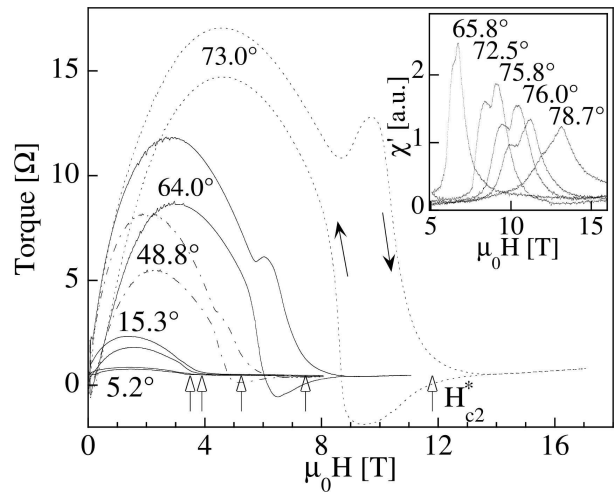}
\caption{Torque versus field at several values of $\theta$  at $T$ =320~mK. Both up and down field sweeps are shown.
Inset: AC susceptibility (real part) versus field for different crystals from the same batch (at $T$=1.2~K). The
excitation field was 38~Oe at a frequency of 72~Hz. } \label{fulltorquefig}
\end{figure}

In Fig.\ \ref{fulltorquefig} we show the measured torque versus field over a large range of field for a small sample
from the same batch as our main crystal (mass=0.4$\mu$g). This small sample was selected for this part of the study to
avoid overstressing the cantilever. The general shape of the curve, a bell shaped bump with peak at around $H_{c2}/2$,
has been explained theoretically in Ref.\ \onlinecite{HaoC91}.   Near $H_{c2}$ there is evidently a pronounced peak
effect which grows in size as the angle is increased.  In addition, there is a significant region above the peak effect
region where the torque is sizeable and remains hysteretic.  Similar features have been reported previously by Angst
\textit{et al.} \cite{AngstPWJKKRK02,AngstPWRKMJKK03} in their torque study of MgB$_2$ single crystals. Their study was
conducted at higher temperature and lower field than the current work, and the peak effect they observed was less
marked.  In the inset to Fig.\ \ref{fulltorquefig} we show data for the ac susceptibility of another crystal from the
same batch (mass=17$\mu$g). Many of the same features evident in the torque data are also visible here, namely the
broad superconducting transition and the pronounced peak effect.

\begin{figure}
\includegraphics*[width=8cm]{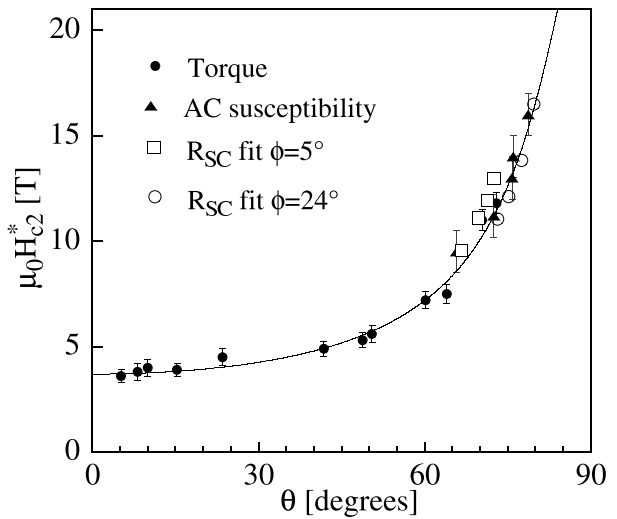}
\caption{$H^*_{c2}$ as determined by torque measurements and AC susceptibility measured at 320mK (solid symbols), along
with a fit to Eq.\ (\ref{gleq}).  The open symbols show the values of $H^*_{c2}(\theta)$ extracted from the fits of
$R_{SC}$ using Eq.\ (\ref{makieq}) for two different values of $\phi$ and are discussed later.} \label{hc2fig}
\end{figure}

The existence of a pronounced curvature in the region of the normal/superconducting
transition along with the peak effect feature means that there is no unambiguous way
of extracting $H_{c2}$ from the measured torque curves. The rounding of the
transition is common in high $T_c$ materials where it is usually attributed to the
presence of thermal or quantum fluctuations, which are enhanced relative to
conventional low $T_c$ materials due to the low dimensionality and high $T_c$ of the
former. MgB$_2$ is not strongly anisotropic, but has a relatively high $H_{c2}$ (for
$H$ parallel to the basal plane) and hence small coherence lengths.  The existence
of hysteresis in this rounding region however, points to an explanation either in
terms of surface superconductivity\cite{RydhWHKKCKKJLKL03} or an as yet unexplained
two gap effect, rather than fluctuations.  It may also be possible that it could
result from crystal inhomogeneity, although this is unlikely as reproducible
behavior is found between crystals from the same batch.

We find that the angular dependence of $H_{c2}$ derived by various extrapolation
schemes (e.g., position of peak effect maximum or a linear extrapolation from low
field) gives different absolute values, but essentially the same form for
$H_{c2}(\theta)/H_{c2}(\theta=0)$ (the same conclusion was found by Angst \textit{et
al.}\cite{AngstPWRKMJKK03}).  As we shall see later, the onset of damping of the
dHvA oscillations essentially coincides with a critical field $H^*_{c2}$ derived by
extrapolating the low field behavior, and so it seems likely that this represents
the bulk upper critical field, at least for the $\pi$ band.

In Fig.\ \ref{hc2fig} we show values of $H^*_{c2}$ extracted from the torque data in
Fig.\ \ref{fulltorquefig}. These values are in good agreement with those
extrapolated from higher temperature studies.\cite{AngstPWJKKRK02} In this figure we
also show $H^*_{c2}$ extracted from ac susceptibility data (inset to Fig.\
\ref{fulltorquefig}) in a similar way.

In an anisotropic superconductor the angular dependence of $H_{c2}$ is usually described by
\begin{equation}
H_{c2}(\theta)=\frac{\gamma_H^2H_{c2}^{\parallel c}}{(\gamma_H^2\cos^2\theta+\sin^2\theta)^\frac{1}{2}}. \label{gleq}
\end{equation}
where $\gamma_H$ is the anisotropy of $H_{c2}$. In MgB$_2$, the contribution of
multiple Fermi surface sheets with different superconducting gaps is known to cause
$\gamma_H$ to increase with decreasing temperature and to cause deviations from the
angular dependence predicted Eq.\ (\ref{gleq}).\cite{AngstPWRKMJKK03} However, these
are small on the scale of Fig.\ \ref{hc2fig}, and are most pronounced near
$\theta=90^\circ$. A fit of this to the data is shown as a solid line in Fig.\
\ref{hc2fig} ($\gamma_H=7\pm0.5$ and $\mu_0H^{\|c}_{c2}= 3.7\pm$~0.3~T ).

\begin{figure}
\includegraphics*[width=8cm]{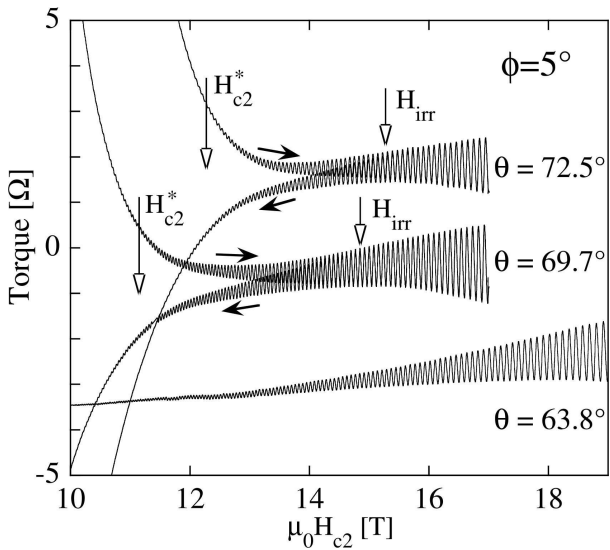}
\caption{Raw torque versus field close to $H_{c2}$ for several angles $\theta$.\label{rawtorquefig}}
\end{figure}

Torque versus field for our main crystal is shown in Fig.\ \ref{rawtorquefig} for three angles ($\theta = 63.8^\circ,
69.7^\circ$ and $72.5^\circ$). At these angles only one dHvA frequency ($F_3$) is visible in our field range. For
$\theta = 69.7^\circ$ pronounced hysteresis is observable which disappears at an irreversibility field $\mu_0H_{\rm
irr}\simeq15.5~$T. There is a small residual field dependence in the background torque which extends to $\sim$16~T. We
note that both these fields are far in excess of our estimate of $H_{c2}$ obtained from the raw torque
($\mu_0H^*_{c2}\simeq11$~T).  Similar behavior can be seen for $\theta=72.5^\circ$.

The field dependent amplitude $A$ of the dHvA oscillations was extracted by fitting $A\sin(2\pi F_3/B+\varphi)+aB+b$
(the linear term accounts for the background torque) to different sections of data comprising of $1\frac{1}{2}$
oscillations. The data were then divided by the weakly field dependent term $B^\frac{3}{2} R_T$ to give
$\tilde{A}\propto R_D R_{SC}$ (see Eq.\ \ref{lkeq}). The quasiparticle effective mass $m^*$ in the expression for $R_T$
was determined by measuring the temperature dependence of the dHvA amplitude [for $F_3$, $m^* = (0.456 \pm 0.005) m_e$
at $\theta = 70.8^\circ$].

\begin{figure}
\includegraphics*[width=8cm]{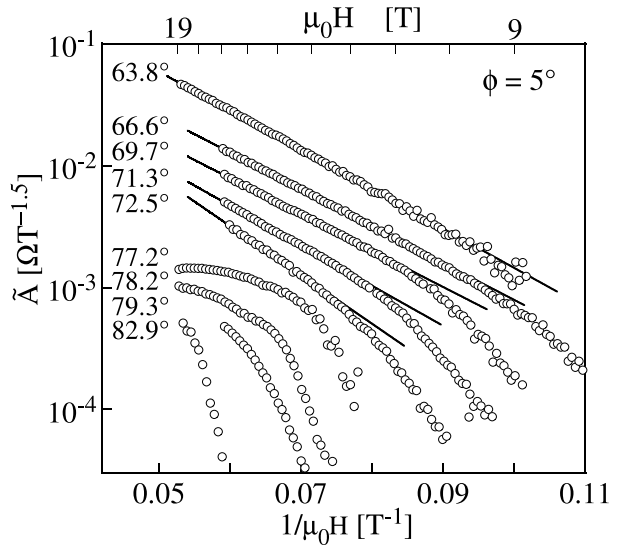} \caption{Amplitude of dHvA oscillations divided by $B^\frac{3}{2} R_T$ versus inverse field, at
several different values of $\theta$ [$\phi=5^\circ$]. The data have been offset for clarity. The actual variation of
amplitude with angle is shown in Fig. \protect\ref{ampthetafig}.} \label{mdinglefig}
\end{figure}

\begin{figure}\center
\includegraphics*[width=8cm]{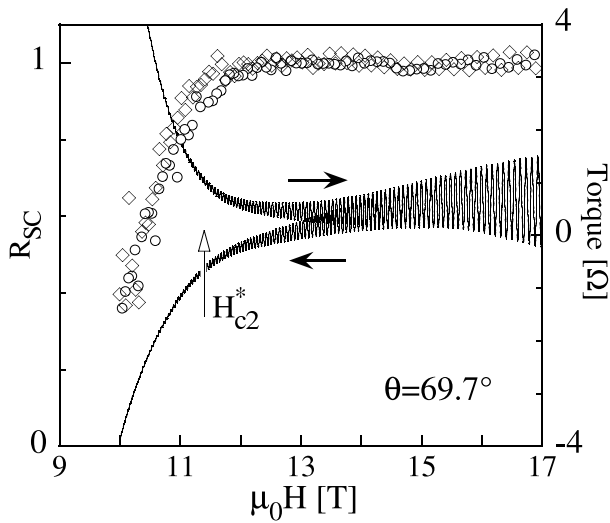}
\caption{Raw torque  and superconducting damping, $R_{SC}$ at $\theta$ = 69.7$^\circ$, for both up and down field sweep
directions (indicated by arrows).} \label{rstorquefig}
\end{figure}

In Fig.\ \ref{mdinglefig} we show $\tilde{A}$ versus inverse field on semi-log axes (`Dingle plot') for several angles.
For $\theta=63.8^\circ$, $\tilde{A}$ varies strictly exponentially with inverse field, $\tilde{A}\propto\exp(-74/B)$
from which we estimate that the quasiparticle mean free path on this orbit is $840\pm20$ \AA. For this angle there is
no evidence of superconductivity in either the background or oscillatory torque (see Fig.\ \ref{rawtorquefig}).  As
$\theta$ is increased $H_{c2}$ increases sharply and the Dingle plots have marked downturns below some critical field.
We attribute this to the opening of the gap as the sample enters the superconducting state. To extract $R_{SC}$ we fit
$\tilde{A}(1/B)$ in the normal state to the exponential expression for $R_D$ and extrapolate this dependence into the
superconducting state (solid line in Fig.\ \ref{mdinglefig}). Dividing $\tilde{A}$ by $R_D$ thus yields $R_{SC}$.  An
example of this is shown in Fig.\ \ref{rstorquefig} in which we also show the raw hysteretic torque.  From this figure
it can be seen that the onset of attenuation of the dHvA signal from $R_{SC}$ does not occur until the field is below
$H_{\rm irr}$, and closely corresponds to the $H^*_{c2}$ value (indicated by an arrow on Fig.\ \ref{rstorquefig})
deduced by extrapolating the lower field data.

Most previous studies of the dHvA effect in the superconducting state have been conducted by the field modulation
technique. In studies where this technique has been used the strong flux pinning close to $H_{c2}$ (peak effect) has
prevented data collection in this region. This did not present a serious problem as the oscillations could be observed
far below $H_{c2}$ [e.g., down to $H_{c2}/5$ in YNi$_2$B$_2$C (Ref.\ \onlinecite{TerashimaHTUAK97}) or $H_{c2}/2$ in
V$_3$Si (Ref. \ \onlinecite{JanssenHHMSW98})]. Studies of YNi$_2$B$_2$C have shown that, in contrast to field
modulation measurements, torque measurements \cite{GollHJJNSWW96} are not affected by this increased pinning. In
MgB$_2$, oscillations are only observable very close to $H_{c2}$ and so all our measurements are essentially carried
out in the peak effect region.

\begin{figure}
\center
\includegraphics*[width=8cm]{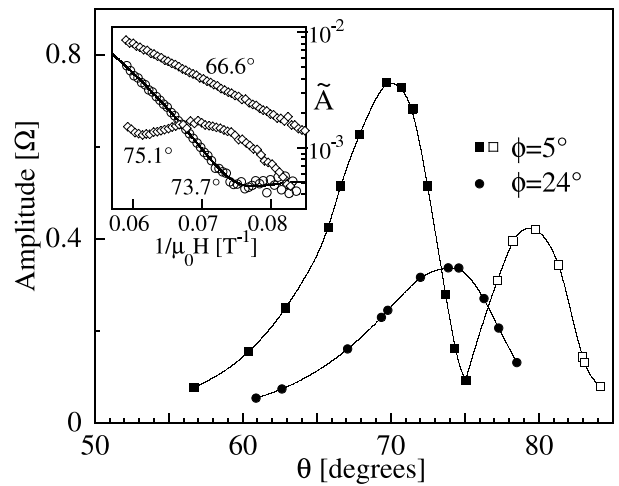}
\caption{Amplitude versus $\theta$ at $\mu _0H=16.5~$T, for two different in-plane rotation angles $\phi=5^\circ$ and
$24^\circ$. Open symbols are the amplitude at 18.5T scaled by the appropriate Dingle factor. Inset: Dingle plots for
$\theta$ near the minimum in amplitude at 75$^\circ$ (Data have been offset for clarity).} \label{ampfnphifig}
\end{figure}

For $\phi$ near to $0^\circ$ there is a pronounced dip is found in the amplitude of the dHvA signal at
$\theta\simeq75^\circ$ (see Fig.\ \ref{ampfnphifig}).  This feature has been present in every crystal we have measured
to date and always occurs when $\theta$ and $\phi$ are adjusted so that $F_3 \simeq 2800$ T.  Previously
\cite{CarringtonMCBHYLYTKK03,YellandCCHMLYT02} this feature has been attributed to a `spin-zero', i.e., an angle where
the angle dependent band mass multiplied by the Stoner factor exactly equals $m_e$/2, so that $R_S=0$. However, a
detailed analysis of the field dependence of the dHvA amplitude for angles above $\theta\simeq 70^\circ$ reveals that
in fact this dip is produced by a beat between two dHvA orbits with very similar frequencies. The inset to Fig.\
\ref{ampfnphifig} shows Dingle plots close to $\theta\simeq75^\circ$.  The data clearly show oscillatory beating damped
by the Dingle factor. A fit that takes account of this effect is shown for $\theta=75.1^\circ$ in Fig.\
\ref{ampfnphifig}. The frequency difference is $\sim 27~$T at $\theta=90^\circ$ and extrapolates to zero at
$\theta=70^\circ$ (for $\phi\sim0^\circ$). The relative amplitudes of the two frequencies are almost equal over the
whole angular range. The characteristics of this orbit are very similar to an additional orbit predicted by Harima.
\cite{Harima02} A full account of this result will be given elsewhere.\cite{carringtonF04}

\begin{figure}
\center
\includegraphics*[width=8cm]{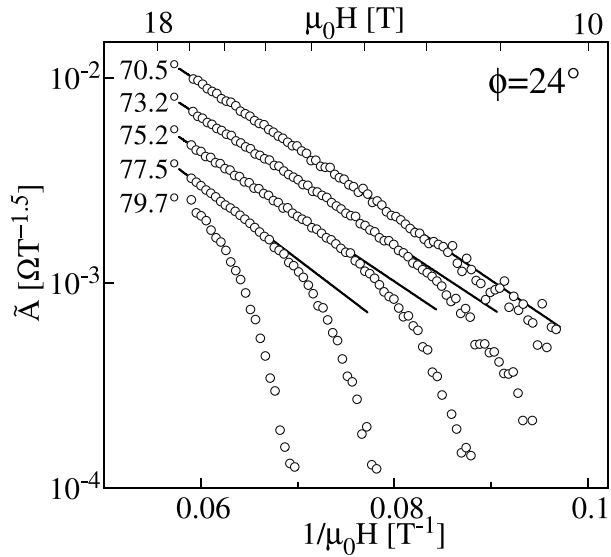}
\caption{Dingle plots for in-plane rotation $\phi=24^\circ$. The solid lines are fits to the Dingle factor $R_D$ [Eq.\
(\protect\ref{lkeq})] for the data above $H_{c2}$.}\label{rdinglefig}
\end{figure}

A consequence of the above dip feature is that at angles close to or beyond $\theta\simeq75^\circ$ it is difficult to
extract $R_{SC}$ as the functional form of $\tilde{A}(1/B)$ is more complicated. For this reason we performed a second
set of sweeps at a different in-plane angle $\phi$. With $\phi=24^\circ$, $F_3(\theta=0)$ is increased to 2930~T and
thus the dip does not occur for any value of $\theta$ (see Fig.\ \ref{ampfnphifig}).  In Fig.\ \ref{rdinglefig} we show
the $\tilde{A}(1/B)$ curves for this in-plane angle.

\begin{figure}
\center
\includegraphics*[width=8cm]{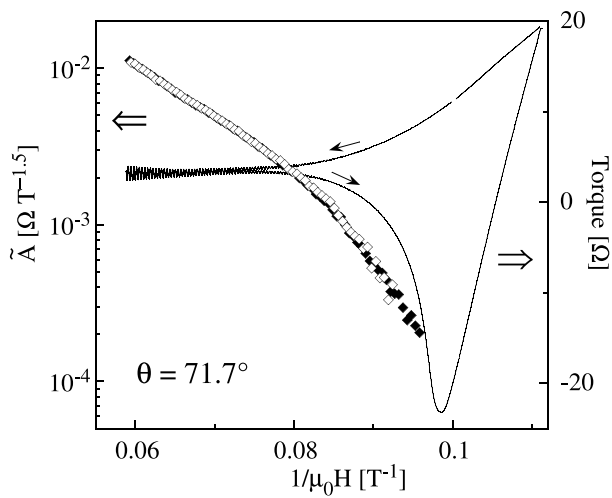}
\caption{Raw torque and $\tilde{A}$ versus inverse field at $\theta=71.7^\circ$ for both up and down field sweeps. The
direction of the field sweep is indicated with the arrows.} \label{updowndinglefig}
\end{figure}

As mentioned earlier, the deflection of the cantilever, either by the background or oscillatory torque, means that the
fields sweeps are not done at strictly constant angle. As the dHvA amplitude is angle dependent this may cause some
additional field dependence to $\tilde{A}$, which becomes large as the sample enters the superconducting state, and
might cause error in our determination of $R_{SC}$. The insensitivity of our results for $R_{SC}$ to this deflection is
perhaps best demonstrated by comparing $\tilde{A}(1/B)$ curves for up and down field sweeps. In Fig.\
\ref{updowndinglefig} we show both the raw torque signal and $\tilde{A}(1/B)$ for $\theta=71.7^\circ$. There is a large
difference in the background torque below $H_{\rm irr}$ in the two cases, which means that the sample is deflected in
opposite directions (maximum deflection $\sim 0.7^\circ$), however it can be see that the drop in $\tilde{A}(1/B)$
below $H_{c2}$ is virtually identical (see also Fig.\ \ref{rstorquefig}).  As a further check we have repeated some our
measurements on a fragment of crystal cut from our main sample which had mass of only 1.3$\mu$g ($\simeq 0.3\times$
mass of main sample). We found identical behavior, showing again that the lever deflection does not affect our results.
The close correspondence of the results for $R_{SC}$ for up and down sweeps is also a strong indication that our data
is not strongly affected by pinning in the peak effect region.

\section{Extracting the gap from the superconducting state damping factor}

dHvA oscillations in the superconducting state have been observed in many different
materials, and in all cases the oscillations persist into the superconducting state
with the same frequency as in the normal state but with reduced amplitude. As the
dHvA amplitude depends exponentially on field, all the materials investigated to
date are those with high $H_{c2}$ values, and are not necessarily conventional. For
example, NbSe$_2$ and Nb$_3$Sn, have recently been suspected of having an exotic gap
structure. \cite{BoakninTPHRHSTSHB03,GuritanuGBWLGJ04} In some ways, MgB$_2$ is the
best understood of all the materials where this phenomena has been observed to date,
although its multiple energy gap structure may complicate the analysis.

Several theoretical models have been proposed to describe the effect (see Ref.
\onlinecite{YasuiK02} and references therein). Two theories have proved most
successful in describing the data to date. Maki\cite{Maki91} used a semiclassical
approach following that of Brandt \textit{et al.}\cite{BrandtPT67}, in which the gap
is approximated by the spatially averaged value of $\Delta^2$ and the magnetic field
is considered to be uniform.  For quasiparticles moving perpendicular to the
magnetic field this model predicts that the excitation spectrum is gapless. It is
this gaplessness which is the physical origin of the quantum oscillations. The extra
damping in the superconducting state is given by
\begin{equation}
R_{SC}=\exp\left[-\pi^\frac{3}{2} \left(\frac{\Delta_E(B)}{\hbar\omega_c}\right)^2
\left(\frac{B}{F}\right)^\frac{1}{2}\right]. \label{makieq}
\end{equation}
In this expression, the \textit{effective} field dependent energy gap $\Delta_E(B)$ is resolved to that on a particular
quasiparticle orbit. An equivalent expression was also derived by different authors starting from alternative physical
pictures. \cite{Stephen92,WassermanS94}

An alternative expression was derived by Miyake,\cite{Miyake93} who considered the
effect of incorporating the zero field BCS quasiparticle energy and occupation
numbers into the usual Lifshitz-Kosevich theory. In the superconducting state the
sharp step in the Fermi function is replaced with the BCS quasiparticle occupation
function ($|u_k|^2$) whose width is set by the superconducting energy gap.  This
gives a more gradual emptying of Landau levels than in the normal state, and hence a
reduced dHvA amplitude.  The size of the damping factor in this model is given by
\begin{equation} R_{SC}=x K_1(x), \quad x=2\pi^2 \frac{\Delta_E(B)}{\hbar \omega_c} \label{miyakeeq}
\end{equation}
where $K_1(x)$ is the Bessel function of the second kind. Miller and Gy\"{o}rffy
\cite{MillerG95} have derived the same result starting from the Bogoliubov-de Gennes
equations.

Studies in other materials have shown that $\Delta_E(B=0)$ extracted by fitting data
to these expressions\cite{JanssenHHMSW98} does not always coincide with the known
energy gap measured by more conventional means (for example tunneling). The
correspondence between $\Delta_E$ and the superconducting gap is different in each
theory, and this may also be different for different materials.  For example,
Janssen \textit{et al.}\cite{JanssenHHMSW98} found that for NbSe$_2$ the Maki model
gave $\Delta_E/\Delta=0.63$ whereas the Miyake model gave, $\Delta_E/\Delta=0.11$.
For V$_3$Si, $\Delta_E/\Delta$ was found to be 2.6 and 0.61 for the two models
respectively. It is generally found that the Maki model gives values for $\Delta_E$
which are much larger than the Miyake model.

To compare these theories with experiment we can take two approaches. The first is to assume a form for the field
dependence of the effective gap and then fit Eq.\ (\ref{makieq}) or Eq.\ (\ref{miyakeeq}) to $R_{SC}(B)$ directly. The
second is to solve Eq.\ (\ref{makieq}) or Eq.\ (\ref{miyakeeq}) for $\Delta_E(B)$ at each field point and then compare
$\Delta_E(B)$ to the expected behavior. We shall show both methods below.

In the usual mean-field BCS theory we expect
\begin{equation}
\frac{\Delta^2_E(B)}{\Delta^2_E(0)} =  1-\frac{B}{B_{c2}}. \label{mfgapeq}
\end{equation}
However, the observed rounding of the superconducting transition means that $d\Delta/dH$ does not change
discontinuously at $H_{c2}$ and so a more appropriate expression is
\begin{equation}
\frac{\Delta^2_E(B)}{\Delta^2_E(0)}=\frac{1}{2}\left[\left(1-\frac{B}{B_{c2}}\right)^2+\alpha^2\right]^\frac{1}{2}+\frac{1}{2}\left(1-\frac{B}{B_{c2}}\right)
\label{flucgapeq}
\end{equation}
This form of $\Delta(B)$ was used by Clayton \textit{et al.}\cite{ClaytonIHMSS02} to
account for the effect of superconducting fluctuations, as it interpolating the mean
field result at low field to the form expected for a strongly fluctuating system
near $H_{c2}$. The phenomenological parameter $\alpha$ sets the strength of the
fluctuations. As it is not clear where the rounding of the superconducting
transition in MgB$_2$ is caused by fluctuations, we will regard $\alpha$ as a
phenomenological parameter which describes the rounding, rather than attributing to
it for now any particular physical significance. In any case, the values of
$\Delta_E(B=0)$ are mostly determined by the portion of the $R_{SC}$ curve well
below $H_{c2}$ (i.e., the data for the higher values of $\theta$).

\begin{figure}
\center
\includegraphics*[width=7cm]{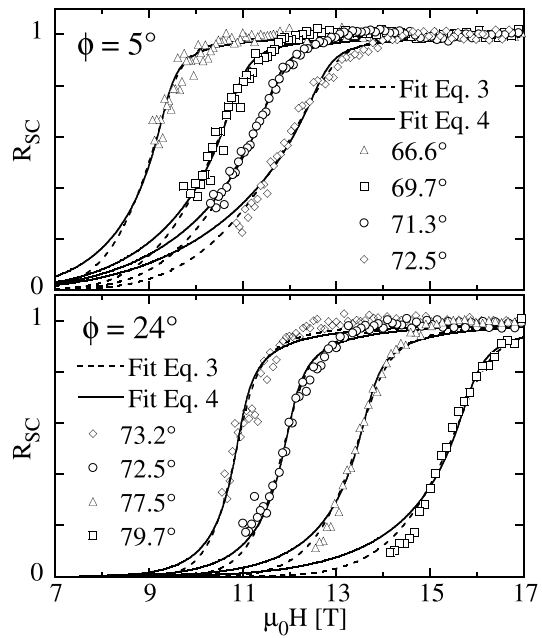}
\caption{$R_{SC}$ for $\phi=5^\circ$(top panel) and $\phi=24^\circ$ (bottom panel). The solid and dashed lines are fits
to the Maki (Eq.\ \ref{makieq}) and Miyaki (Eq.\ \ref{miyakeeq}) theories for the superconducting state
damping}\label{rsfig}
\end{figure}

Combining Eq.\ \ref{flucgapeq} with either Eq.\ (\ref{makieq}) or Eq.\ (\ref{miyakeeq}) results in 3 fitting parameters
per curve ($\Delta_E(0)$, $H_{c2}$ and $\alpha$). We found that in general there was too much covariance between the
parameters to arrive at accurate values for $\Delta_E(0)$ if we allowed all three to vary between the fits at each
angle. As we do not expect either $\Delta_E(0)$ or $\alpha$ to vary strongly with $\theta$ we found that more
consistent results were obtained by fitting to the data for $R_{SC}(B)$ at all angles simultaneously, only allowing
$H_{c2}$ to vary as a function of angle.

\begin{table}
\caption{Parameters extracted from the fits to $R_{SC}$}
\begin{center}
\begin{tabular}{lcccc}
Theory&$\phi$ &$\Delta_E(B=0)$&$10^3 \times \alpha$\\
\hline\hline
Maki&5$^\circ$&200$\pm40$~K&55$\pm10$\\
Maki&24$^\circ$&320$\pm40$~K&34$\pm10$\\
Miyake&5$^\circ$&32$\pm10$~K&27$\pm5$\\
Miyake&24$^\circ$&60$\pm10$~K&22$\pm5$\\
\hline\hline
\end{tabular}
\end{center}
\label{fitparamstbl}
\end{table}

The fits obtained for both $\phi=5^\circ$ and $\phi=24^\circ$ are shown in Fig.
\ref{rsfig}. In these fits we have taken $m_B(\theta)=0.315m_eF_3/F_3^0$, i.e.,
assuming the band mass scales like the dHvA frequency.\cite{MazinK02,massnote} The
values of $\Delta_E(B=0)$ and $\alpha$ and are shown in Table \ref{fitparamstbl}. It
can be seen that both theories provide an adequate fit to the experimental data. The
main difference is in the size of the extracted superconducting gap.  As orbit 3 is
on the electron like $\pi$ sheet of Fermi surface we should compare these values
with the zero field value for the smaller MgB$_2$ gap ($\Delta^s\simeq$ 29~K).
\cite{ManzanoCHLYT02} The Miyake fit gives values comparable to this whereas the
Maki fit gives values up to 10 times larger. The trend is similar to that found for
NbSe$_2$. The values of $\alpha$ are also somewhat larger in the Maki fits to those
in the Miyake fits. The values of $H_{c2}$ extracted from our fits are shown in
Fig.\ \ref{hc2fig}. The values are very similar to those extracted from the
background torque, consistent with our interpretation of our extrapolated $H^*_{c2}$
as the bulk critical field for the $\pi$ band. There is a consistent difference in
$H_{c2}(\theta)$ values for the two values of $\phi$, with the values for
$\phi=5^\circ$ being $\sim 10$\% higher than those for $\phi=24^\circ$.  This
implies a slight in-plane anisotropy of $H_{c2}$.

\begin{figure}
\center
\includegraphics*[width=8cm]{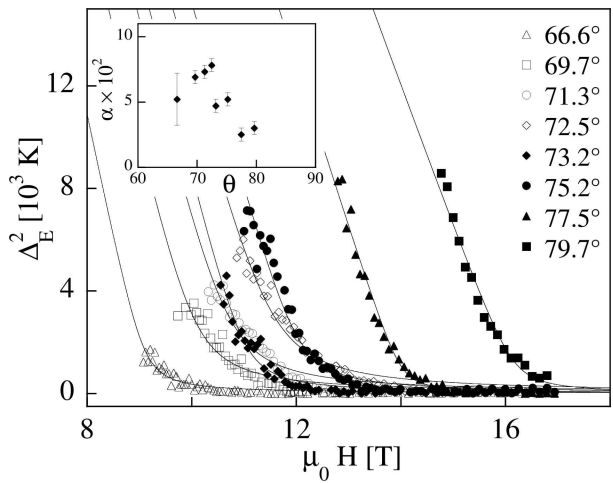} \caption{Calculation of the field dependence of the superconducting gap, by inverting the
$R_{SC}$ curves in Fig.\ \protect\ref{rsfig}  using the Maki [Eq.\ (\protect\ref{makieq})] model. The solid lines are
fits to the field dependent gap [Eq.\ (\protect\ref{flucgapeq})](see text). The inset shows the variation of the fit
parameter $\alpha$ with $\theta$.} \label{rgapfig}
\end{figure}

Although there is some ambiguity between the theories as to the magnitude of the
gap, it is clear that a sizeable gap does exist on this sheet of Fermi surface right
up to $H_{c2}$. There is some difference between the gap values derived from the
data at the two in-plane rotation angles. We note that although there is a large
difference between the gaps on the $\sigma$ and $\pi$ Fermi surface sheets, the
difference within a sheet (or indeed pair of sheets) is expected to be
small.\cite{MazinAJGDK04} One reason for this orientation dependence is that the
data for $\phi=24^\circ$ cover a wider range of $H/H_{c2}$ and hence, the
extrapolated gap value is more accurate. For the runs with different $\phi$ where
the $H_{c2}$ values are close, the $R_{SC}$ curves are very similar, and any
difference results from the field range of the fit. Next, we show in Fig.\
\ref{rgapfig} the field dependent effective gap calculated by inverting Eq.\
(\ref{makieq}), along with fits to the field dependent gap given by Eq.\
(\ref{flucgapeq}).  The main difference between using Eq.\ (\ref{miyakeeq}) rather
than Eq.\ (\ref{makieq}) is just in the magnitude of the extracted $\Delta_E$ rather
than its field dependence. In the fits we have fixed $\Delta_E(B=0)$ to 300~K for
all values of $\theta$ and $\phi$ and allowed $\alpha$ to vary. The curvature near
$H_{c2}$ and the approximately linear behavior of $\Delta_E^2(B)$ at lower field are
evident. The covariance between the gap and $H_{c2}$ mentioned above arises from the
rounding at the transition and the small field range available at some angles. There
is however a discernible progressive change in shape of the $\Delta_E(B)$ curves as
$\theta$ and $\phi$ are varied. This is reflected in the angular dependence of
$\alpha$ in the inset to the figure.  This may be an artifact of the data analysis
but could also result from a two gap effect.

It is clear that the extrapolation to the zero field gap is not a trivial one as our data does not extend below
0.85$H_{c2}$. This is particularly significant in a material like MgB$_2$, where non mean-field gap behavior is
expected.  Numerical calculations\cite{GraserDS04,dahmprivate} of the field dependence of the energy gaps in MgB$_2$
show that they differ considerably from the mean field behavior given by Eq.\ (\ref{mfgapeq}). Close to $H_{c2}$, for
$H\|ab$, we find that the numerical data can be approximated by
$\Delta^2_\pi(B)=0.24\Delta^2_\pi(0)(1-H/H_{c2})^{1.4}$. This accounts naturally for some (but not all) of the rounding
in our $R_{SC}$ plots. The steep drop of $\Delta^2_\pi$ at low field found in the calculations means that a linear
extrapolation from high field underestimates the true zero field gap by approximately 50\%.

Recently, there has been further theoretical work on the dHvA effect in superconductors.   Duncan and Gy\"{o}rffy
\cite{DuncanG03} have extended the work in Ref.\ \onlinecite{MillerG95} and given a new formula for $R_{SC}$.  We have
fitted our data to this formula and find fits that are comparable in quality to those in Fig.\ \ref{rsfig}, with
$\Delta_E(B=0)=200\pm20$~K, for $\phi=5^\circ$. This is very close to the values calculated from the Maki formula [Eq.\
(\ref{makieq})]. Yasui and Kita \cite{YasuiK02} have made a detailed numerical investigation of the approximations used
in many of the other previous superconducting state dHvA theories. They proposed an equation for $R_{SC}$ which should
give quantitative values for the energy gap which can be compared to the \textit{actual} thermodynamic values.  Their
expression for $R_{SC}$ is
\begin{equation}
R_{SC}= \exp\left[-2\pi\beta\left(\frac{\Delta_E(B)}{\hbar\omega_c}\right)^2\right] \label{ykeq}
\end{equation}
where $\beta=0.0625$ is a numerically evaluated constant.  This was shown to correctly describe the damping in Nb$_3$Sn
and NbSe$_2$ with $\Delta_E$ approximately equal to the actual superconducting energy gap. It can be seen that this is
almost exactly the same as Maki's formula, Eq.\ (\ref{makieq}), except for a factor
$2\beta/[\pi^\frac{1}{2}(F/B)^\frac{1}{2}]$ which equals 0.94 for $F=2700$~T and $B=15$~T.  Hence, the gap values
extracted by fitting this formula are virtually identical to those from Eq.\ (\ref{makieq}) i.e., up to 10 times larger
then the gap values extracted by other means.

\section{Conclusions}
In conclusion, we have made a detailed study of the attenuation of the dHvA signal in MgB$_2$ as it enters the
superconducting state. Only a single orbit on the electron-like $\pi$ band is observable. The data clearly show that a
sizeable gap exists on this orbit even at high field.  The transition from normal to superconducting states as seen in
both the background magnetisation as well as the dHvA amplitude are rather broad, which may result from either
fluctuation effects or a two-gap effect.  The data provide a new test for theories of the dHvA effect in the
superconducting state being applied for the first time to a material which clearly has two superconducting gaps.

Both the Miyake and the Maki theories fit the data well, although the Miyake theory produces gap values which are much
closer to those expected from the low field data for the $\pi$ band.  The most recent theoretical work
\cite{YasuiK02,DuncanG03} however, suggests that the Miyake theory should seriously overestimate the damping and that
the Maki model should provide an accurate quantitative estimate of the average gap on the dHvA orbit. Our data is in
serious disagreement with this as the gap values found from the Maki model fits are more than a factor 10 times larger
than the expected zero field $\pi$ band gap.  We conclude that the present theories of the dHvA effect in the
superconducting state, when applied to MgB$_2$, dramatically underestimate the damping.

\section{Acknowledgments}
We would like to thank J.R.\ Cooper, S.M.\ Hayden, B. Gy\"{o}rffy and P.J.\ Meeson for useful discussions.


\begin{thebibliography}{58}
\expandafter\ifx\csname natexlab\endcsname\relax\def\natexlab#1{#1}\fi \expandafter\ifx\csname
bibnamefont\endcsname\relax
  \def\bibnamefont#1{#1}\fi
\expandafter\ifx\csname bibfnamefont\endcsname\relax
  \def\bibfnamefont#1{#1}\fi
\expandafter\ifx\csname citenamefont\endcsname\relax
  \def\citenamefont#1{#1}\fi
\expandafter\ifx\csname url\endcsname\relax
  \def\url#1{\texttt{#1}}\fi
\expandafter\ifx\csname urlprefix\endcsname\relax\def\urlprefix{URL }\fi \providecommand{\bibinfo}[2]{#2}
\providecommand{\eprint}[2][]{\url{#2}}

\bibitem[{\citenamefont{Iavarone et~al.}(2002)\citenamefont{Iavarone,
  Karapetrov, Koshelev, Kwok, Crabtree, Hinks, Kang, Choi, Kim, Kim, and
  Lee}}]{IavaroneKKKCHKCKKL02}
\bibinfo{author}{\bibfnamefont{M.}~\bibnamefont{Iavarone}},
  \bibinfo{author}{\bibfnamefont{G.}~\bibnamefont{Karapetrov}},
  \bibinfo{author}{\bibfnamefont{A.~E.} \bibnamefont{Koshelev}},
  \bibinfo{author}{\bibfnamefont{W.~K.} \bibnamefont{Kwok}},
  \bibinfo{author}{\bibfnamefont{G.~W.} \bibnamefont{Crabtree}},
  \bibinfo{author}{\bibfnamefont{D.~G.} \bibnamefont{Hinks}},
  \bibinfo{author}{\bibfnamefont{W.~N.} \bibnamefont{Kang}},
  \bibinfo{author}{\bibfnamefont{E.~M.} \bibnamefont{Choi}},
  \bibinfo{author}{\bibfnamefont{H.~J.} \bibnamefont{Kim}},
  \bibinfo{author}{\bibfnamefont{H.~J.} \bibnamefont{Kim}}, \bibnamefont{and}
  \bibinfo{author}{\bibfnamefont{S.~I.} \bibnamefont{Lee}},
  \bibinfo{journal}{Phys. Rev. Lett.} \textbf{\bibinfo{volume}{89}},
  \bibinfo{pages}{187002} (\bibinfo{year}{2002}).

\bibitem[{\citenamefont{Bouquet et~al.}(2001)\citenamefont{Bouquet, Wang,
  Fisher, Hinks, Jorgensen, Junod, and Phillips}}]{BouquetWFHJJP01}
\bibinfo{author}{\bibfnamefont{F.}~\bibnamefont{Bouquet}},
  \bibinfo{author}{\bibfnamefont{Y.}~\bibnamefont{Wang}},
  \bibinfo{author}{\bibfnamefont{R.~A.} \bibnamefont{Fisher}},
  \bibinfo{author}{\bibfnamefont{D.~G.} \bibnamefont{Hinks}},
  \bibinfo{author}{\bibfnamefont{J.~D.} \bibnamefont{Jorgensen}},
  \bibinfo{author}{\bibfnamefont{A.}~\bibnamefont{Junod}}, \bibnamefont{and}
  \bibinfo{author}{\bibfnamefont{N.~E.} \bibnamefont{Phillips}},
  \bibinfo{journal}{Europhys. Lett.} \textbf{\bibinfo{volume}{56}},
  \bibinfo{pages}{856} (\bibinfo{year}{2001}).

\bibitem[{\citenamefont{Bouquet et~al.}(2002)\citenamefont{Bouquet, Wang,
  Sheikin, Plackowski, Junod, Lee, and Tajima}}]{BouquetWSPJLT02}
\bibinfo{author}{\bibfnamefont{F.}~\bibnamefont{Bouquet}},
  \bibinfo{author}{\bibfnamefont{Y.}~\bibnamefont{Wang}},
  \bibinfo{author}{\bibfnamefont{I.}~\bibnamefont{Sheikin}},
  \bibinfo{author}{\bibfnamefont{T.}~\bibnamefont{Plackowski}},
  \bibinfo{author}{\bibfnamefont{A.}~\bibnamefont{Junod}},
  \bibinfo{author}{\bibfnamefont{S.}~\bibnamefont{Lee}}, \bibnamefont{and}
  \bibinfo{author}{\bibfnamefont{S.}~\bibnamefont{Tajima}},
  \bibinfo{journal}{Phys. Rev. Lett.} \textbf{\bibinfo{volume}{89}},
  \bibinfo{pages}{257001} (\bibinfo{year}{2002}).

\bibitem[{\citenamefont{Manzano et~al.}(2002)\citenamefont{Manzano, Carrington,
  Hussey, Lee, Yamamoto, and Tajima}}]{ManzanoCHLYT02}
\bibinfo{author}{\bibfnamefont{F.}~\bibnamefont{Manzano}},
  \bibinfo{author}{\bibfnamefont{A.}~\bibnamefont{Carrington}},
  \bibinfo{author}{\bibfnamefont{N.~E.} \bibnamefont{Hussey}},
  \bibinfo{author}{\bibfnamefont{S.}~\bibnamefont{Lee}},
  \bibinfo{author}{\bibfnamefont{A.}~\bibnamefont{Yamamoto}}, \bibnamefont{and}
  \bibinfo{author}{\bibfnamefont{S.}~\bibnamefont{Tajima}},
  \bibinfo{journal}{Phys. Rev. Lett.} \textbf{\bibinfo{volume}{88}},
  \bibinfo{pages}{047002} (\bibinfo{year}{2002}).

\bibitem[{\citenamefont{Souma et~al.}(2003)\citenamefont{Souma, Machida, Sato,
  Takahashi, Matsui, Wang, Ding, Kaminski, Campuzano, Sasaki, and
  Kadowaki}}]{SoumaMSTMWDKCSK03}
\bibinfo{author}{\bibfnamefont{S.}~\bibnamefont{Souma}},
  \bibinfo{author}{\bibfnamefont{Y.}~\bibnamefont{Machida}},
  \bibinfo{author}{\bibfnamefont{T.}~\bibnamefont{Sato}},
  \bibinfo{author}{\bibfnamefont{T.}~\bibnamefont{Takahashi}},
  \bibinfo{author}{\bibfnamefont{H.}~\bibnamefont{Matsui}},
  \bibinfo{author}{\bibfnamefont{S.~C.} \bibnamefont{Wang}},
  \bibinfo{author}{\bibfnamefont{H.}~\bibnamefont{Ding}},
  \bibinfo{author}{\bibfnamefont{A.}~\bibnamefont{Kaminski}},
  \bibinfo{author}{\bibfnamefont{J.~C.} \bibnamefont{Campuzano}},
  \bibinfo{author}{\bibfnamefont{S.}~\bibnamefont{Sasaki}}, \bibnamefont{and}
  \bibinfo{author}{\bibfnamefont{K.}~\bibnamefont{Kadowaki}},
  \bibinfo{journal}{Nature} \textbf{\bibinfo{volume}{423}}, \bibinfo{pages}{65}
  (\bibinfo{year}{2003}).

\bibitem[{\citenamefont{Tsuda et~al.}(2003)\citenamefont{Tsuda, Yokoya, Takano,
  Kito, Matsushita, Yin, Itoh, Harima, and Shin}}]{TsudaYTKMYIHS03}
\bibinfo{author}{\bibfnamefont{S.}~\bibnamefont{Tsuda}},
  \bibinfo{author}{\bibfnamefont{T.}~\bibnamefont{Yokoya}},
  \bibinfo{author}{\bibfnamefont{Y.}~\bibnamefont{Takano}},
  \bibinfo{author}{\bibfnamefont{H.}~\bibnamefont{Kito}},
  \bibinfo{author}{\bibfnamefont{A.}~\bibnamefont{Matsushita}},
  \bibinfo{author}{\bibfnamefont{F.}~\bibnamefont{Yin}},
  \bibinfo{author}{\bibfnamefont{J.}~\bibnamefont{Itoh}},
  \bibinfo{author}{\bibfnamefont{H.}~\bibnamefont{Harima}}, \bibnamefont{and}
  \bibinfo{author}{\bibfnamefont{S.}~\bibnamefont{Shin}},
  \bibinfo{journal}{Phys. Rev. Lett.} \textbf{\bibinfo{volume}{91}},
  \bibinfo{pages}{127001} (\bibinfo{year}{2003}).

\bibitem[{\citenamefont{Shen et~al.}(1965)\citenamefont{Shen, Senozan, and
  Phillips}}]{ShenSP65}
\bibinfo{author}{\bibfnamefont{L.~Y.~L.} \bibnamefont{Shen}},
  \bibinfo{author}{\bibfnamefont{N.~M.} \bibnamefont{Senozan}},
  \bibnamefont{and} \bibinfo{author}{\bibfnamefont{N.~E.}
  \bibnamefont{Phillips}}, \bibinfo{journal}{Phys. Rev. Lett.}
  \textbf{\bibinfo{volume}{14}}, \bibinfo{pages}{1025} (\bibinfo{year}{1965}).

\bibitem[{\citenamefont{Binnig et~al.}(1980)\citenamefont{Binnig, Baratoff,
  Hoenig, and Bednorz}}]{BinnigBHB80}
\bibinfo{author}{\bibfnamefont{G.}~\bibnamefont{Binnig}},
  \bibinfo{author}{\bibfnamefont{A.}~\bibnamefont{Baratoff}},
  \bibinfo{author}{\bibfnamefont{H.~E.} \bibnamefont{Hoenig}},
  \bibnamefont{and} \bibinfo{author}{\bibfnamefont{J.~G.}
  \bibnamefont{Bednorz}}, \bibinfo{journal}{Phys. Rev. Lett.}
  \textbf{\bibinfo{volume}{45}}, \bibinfo{pages}{1352} (\bibinfo{year}{1980}).

\bibitem[{\citenamefont{Kortus et~al.}(2001)\citenamefont{Kortus, Mazin,
  Belashchenko, Antropov, and Boyer}}]{KortusMBAB01}
\bibinfo{author}{\bibfnamefont{J.}~\bibnamefont{Kortus}},
  \bibinfo{author}{\bibfnamefont{I.~I.} \bibnamefont{Mazin}},
  \bibinfo{author}{\bibfnamefont{K.~D.} \bibnamefont{Belashchenko}},
  \bibinfo{author}{\bibfnamefont{V.~P.} \bibnamefont{Antropov}},
  \bibnamefont{and} \bibinfo{author}{\bibfnamefont{L.~L.} \bibnamefont{Boyer}},
  \bibinfo{journal}{Phys. Rev. Lett.} \textbf{\bibinfo{volume}{86}},
  \bibinfo{pages}{4656} (\bibinfo{year}{2001}).

\bibitem[{\citenamefont{Liu et~al.}(2001)\citenamefont{Liu, Mazin, and
  Kortus}}]{LiuMK01}
\bibinfo{author}{\bibfnamefont{A.~Y.} \bibnamefont{Liu}},
  \bibinfo{author}{\bibfnamefont{I.~I.} \bibnamefont{Mazin}}, \bibnamefont{and}
  \bibinfo{author}{\bibfnamefont{J.}~\bibnamefont{Kortus}},
  \bibinfo{journal}{Phys. Rev. Lett.} \textbf{\bibinfo{volume}{8708}},
  \bibinfo{pages}{087005} (\bibinfo{year}{2001}).

\bibitem[{\citenamefont{Choi et~al.}(2002)\citenamefont{Choi, Roundy, Sun,
  Cohen, and Louie}}]{ChoiRSCL02}
\bibinfo{author}{\bibfnamefont{H.~J.} \bibnamefont{Choi}},
  \bibinfo{author}{\bibfnamefont{D.}~\bibnamefont{Roundy}},
  \bibinfo{author}{\bibfnamefont{H.}~\bibnamefont{Sun}},
  \bibinfo{author}{\bibfnamefont{M.~L.} \bibnamefont{Cohen}}, \bibnamefont{and}
  \bibinfo{author}{\bibfnamefont{S.~G.} \bibnamefont{Louie}},
  \bibinfo{journal}{Nature} \textbf{\bibinfo{volume}{418}},
  \bibinfo{pages}{758} (\bibinfo{year}{2002}).

\bibitem[{\citenamefont{Carrington and Manzano}(2003)}]{CarringtonM03}
\bibinfo{author}{\bibfnamefont{A.}~\bibnamefont{Carrington}} \bibnamefont{and}
  \bibinfo{author}{\bibfnamefont{F.}~\bibnamefont{Manzano}},
  \bibinfo{journal}{Physica C} \textbf{\bibinfo{volume}{385}},
  \bibinfo{pages}{205} (\bibinfo{year}{2003}).

\bibitem[{\citenamefont{Gonnelli et~al.}(2002)\citenamefont{Gonnelli, Daghero,
  Ummarino, Stepanov, Jun, Kazakov, and Karpinski}}]{GonnelliDUSJKK02}
\bibinfo{author}{\bibfnamefont{R.~S.} \bibnamefont{Gonnelli}},
  \bibinfo{author}{\bibfnamefont{D.}~\bibnamefont{Daghero}},
  \bibinfo{author}{\bibfnamefont{G.~A.} \bibnamefont{Ummarino}},
  \bibinfo{author}{\bibfnamefont{V.~A.} \bibnamefont{Stepanov}},
  \bibinfo{author}{\bibfnamefont{J.}~\bibnamefont{Jun}},
  \bibinfo{author}{\bibfnamefont{S.~M.} \bibnamefont{Kazakov}},
  \bibnamefont{and}
  \bibinfo{author}{\bibfnamefont{J.}~\bibnamefont{Karpinski}},
  \bibinfo{journal}{Phys. Rev. Lett.} \textbf{\bibinfo{volume}{89}},
  \bibinfo{pages}{247004} (\bibinfo{year}{2002}).

\bibitem[{\citenamefont{Szabo et~al.}(2001)\citenamefont{Szabo, Samuely,
  Kacmarcik, Klein, Marcus, Fruchart, Miraglia, Marcenat, and
  Jansen}}]{SzaboSKKMFMMJ01}
\bibinfo{author}{\bibfnamefont{P.}~\bibnamefont{Szabo}},
  \bibinfo{author}{\bibfnamefont{P.}~\bibnamefont{Samuely}},
  \bibinfo{author}{\bibfnamefont{J.}~\bibnamefont{Kacmarcik}},
  \bibinfo{author}{\bibfnamefont{T.}~\bibnamefont{Klein}},
  \bibinfo{author}{\bibfnamefont{J.}~\bibnamefont{Marcus}},
  \bibinfo{author}{\bibfnamefont{D.}~\bibnamefont{Fruchart}},
  \bibinfo{author}{\bibfnamefont{S.}~\bibnamefont{Miraglia}},
  \bibinfo{author}{\bibfnamefont{C.}~\bibnamefont{Marcenat}}, \bibnamefont{and}
  \bibinfo{author}{\bibfnamefont{A.~G.~M.} \bibnamefont{Jansen}},
  \bibinfo{journal}{Phys. Rev. Lett.} \textbf{\bibinfo{volume}{87}},
  \bibinfo{pages}{137005} (\bibinfo{year}{2001}).

\bibitem[{\citenamefont{Graser et~al.}(2004)\citenamefont{Graser, Dahm, and
  Schopohl}}]{GraserDS04}
\bibinfo{author}{\bibfnamefont{S.}~\bibnamefont{Graser}},
  \bibinfo{author}{\bibfnamefont{T.}~\bibnamefont{Dahm}}, \bibnamefont{and}
  \bibinfo{author}{\bibfnamefont{N.}~\bibnamefont{Schopohl}},
  \bibinfo{journal}{Phys. Rev. B} \textbf{\bibinfo{volume}{69}},
  \bibinfo{pages}{014511} (\bibinfo{year}{2004}).

\bibitem[{\citenamefont{Dahm}()}]{dahmprivate}
\bibinfo{author}{\bibfnamefont{T.}~\bibnamefont{Dahm}},
  \bibinfo{howpublished}{(Private communication)}.

\bibitem[{\citenamefont{Daghero et~al.}(2003)\citenamefont{Daghero, Gonnelli,
  Ummarino, Stepanov, Jun, Kazakov, and Karpinski}}]{DagheroGUSJKK03}
\bibinfo{author}{\bibfnamefont{D.}~\bibnamefont{Daghero}},
  \bibinfo{author}{\bibfnamefont{R.~S.} \bibnamefont{Gonnelli}},
  \bibinfo{author}{\bibfnamefont{G.~A.} \bibnamefont{Ummarino}},
  \bibinfo{author}{\bibfnamefont{V.~A.} \bibnamefont{Stepanov}},
  \bibinfo{author}{\bibfnamefont{J.}~\bibnamefont{Jun}},
  \bibinfo{author}{\bibfnamefont{S.~M.} \bibnamefont{Kazakov}},
  \bibnamefont{and}
  \bibinfo{author}{\bibfnamefont{J.}~\bibnamefont{Karpinski}},
  \bibinfo{journal}{Physica C} \textbf{\bibinfo{volume}{385}},
  \bibinfo{pages}{255} (\bibinfo{year}{2003}).

\bibitem[{\citenamefont{Bugoslavsky et~al.}(2003)\citenamefont{Bugoslavsky,
  Miyoshi, Perkins, Caplin, Cohen, Pogrebnyakov, and Xi}}]{BugoslavskyMPCCPX03}
\bibinfo{author}{\bibfnamefont{Y.}~\bibnamefont{Bugoslavsky}},
  \bibinfo{author}{\bibfnamefont{Y.}~\bibnamefont{Miyoshi}},
  \bibinfo{author}{\bibfnamefont{G.}~\bibnamefont{Perkins}},
  \bibinfo{author}{\bibfnamefont{A.}~\bibnamefont{Caplin}},
  \bibinfo{author}{\bibfnamefont{L.}~\bibnamefont{Cohen}},
  \bibinfo{author}{\bibfnamefont{A.}~\bibnamefont{Pogrebnyakov}},
  \bibnamefont{and} \bibinfo{author}{\bibfnamefont{X.}~\bibnamefont{Xi}}
  (\bibinfo{year}{2003}), \bibinfo{note}{cond-mat/0307540}.

\bibitem[{\citenamefont{Gonnelli et~al.}(2003)\citenamefont{Gonnelli, Daghero,
  Calzolari, Ummarino, Dellarocca, Stepanov, Jun, Kazakov, and
  Karpinski}}]{Gonnelli03}
\bibinfo{author}{\bibfnamefont{R.}~\bibnamefont{Gonnelli}},
  \bibinfo{author}{\bibfnamefont{D.}~\bibnamefont{Daghero}},
  \bibinfo{author}{\bibfnamefont{A.}~\bibnamefont{Calzolari}},
  \bibinfo{author}{\bibfnamefont{G.}~\bibnamefont{Ummarino}},
  \bibinfo{author}{\bibfnamefont{V.}~\bibnamefont{Dellarocca}},
  \bibinfo{author}{\bibfnamefont{V.}~\bibnamefont{Stepanov}},
  \bibinfo{author}{\bibfnamefont{J.}~\bibnamefont{Jun}},
  \bibinfo{author}{\bibfnamefont{S.}~\bibnamefont{Kazakov}}, \bibnamefont{and}
  \bibinfo{author}{\bibfnamefont{J.}~\bibnamefont{Karpinski}}
  (\bibinfo{year}{2003}), \bibinfo{note}{cond-mat/0308152}.

\bibitem[{\citenamefont{Corcoran
  et~al.}(1994{\natexlab{a}})\citenamefont{Corcoran, Meeson, Onuki, Probst,
  Springford, Takita, Harima, Guo, and Gyorffy}}]{CorcoranMOPSTHGG94}
\bibinfo{author}{\bibfnamefont{R.}~\bibnamefont{Corcoran}},
  \bibinfo{author}{\bibfnamefont{P.}~\bibnamefont{Meeson}},
  \bibinfo{author}{\bibfnamefont{Y.}~\bibnamefont{Onuki}},
  \bibinfo{author}{\bibfnamefont{P.~A.} \bibnamefont{Probst}},
  \bibinfo{author}{\bibfnamefont{M.}~\bibnamefont{Springford}},
  \bibinfo{author}{\bibfnamefont{K.}~\bibnamefont{Takita}},
  \bibinfo{author}{\bibfnamefont{H.}~\bibnamefont{Harima}},
  \bibinfo{author}{\bibfnamefont{G.~Y.} \bibnamefont{Guo}}, \bibnamefont{and}
  \bibinfo{author}{\bibfnamefont{B.~L.} \bibnamefont{Gyorffy}},
  \bibinfo{journal}{J. Phys.-Condes. Matter} \textbf{\bibinfo{volume}{6}},
  \bibinfo{pages}{4479} (\bibinfo{year}{1994}{\natexlab{a}}).

\bibitem[{\citenamefont{Corcoran
  et~al.}(1994{\natexlab{b}})\citenamefont{Corcoran, Harrison, Hayden, Meeson,
  Springford, and Vanderwel}}]{CorcoranHHMSV94}
\bibinfo{author}{\bibfnamefont{R.}~\bibnamefont{Corcoran}},
  \bibinfo{author}{\bibfnamefont{N.}~\bibnamefont{Harrison}},
  \bibinfo{author}{\bibfnamefont{S.~M.} \bibnamefont{Hayden}},
  \bibinfo{author}{\bibfnamefont{P.}~\bibnamefont{Meeson}},
  \bibinfo{author}{\bibfnamefont{M.}~\bibnamefont{Springford}},
  \bibnamefont{and} \bibinfo{author}{\bibfnamefont{P.~J.}
  \bibnamefont{Vanderwel}}, \bibinfo{journal}{Phys. Rev. Lett.}
  \textbf{\bibinfo{volume}{72}}, \bibinfo{pages}{701}
  (\bibinfo{year}{1994}{\natexlab{b}}).

\bibitem[{\citenamefont{Harrison et~al.}(1994)\citenamefont{Harrison, Hayden,
  Meeson, Springford, Vanderwel, and Menovsky}}]{HarrisonHMSVM94}
\bibinfo{author}{\bibfnamefont{N.}~\bibnamefont{Harrison}},
  \bibinfo{author}{\bibfnamefont{S.~M.} \bibnamefont{Hayden}},
  \bibinfo{author}{\bibfnamefont{P.}~\bibnamefont{Meeson}},
  \bibinfo{author}{\bibfnamefont{M.}~\bibnamefont{Springford}},
  \bibinfo{author}{\bibfnamefont{P.~J.} \bibnamefont{Vanderwel}},
  \bibnamefont{and} \bibinfo{author}{\bibfnamefont{A.~A.}
  \bibnamefont{Menovsky}}, \bibinfo{journal}{Phys. Rev. B}
  \textbf{\bibinfo{volume}{50}}, \bibinfo{pages}{4208} (\bibinfo{year}{1994}).

\bibitem[{\citenamefont{Clayton et~al.}(2002)\citenamefont{Clayton, Ito,
  Hayden, Meeson, Springford, and Saito}}]{ClaytonIHMSS02}
\bibinfo{author}{\bibfnamefont{N.~J.} \bibnamefont{Clayton}},
  \bibinfo{author}{\bibfnamefont{H.}~\bibnamefont{Ito}},
  \bibinfo{author}{\bibfnamefont{S.~M.} \bibnamefont{Hayden}},
  \bibinfo{author}{\bibfnamefont{P.~J.} \bibnamefont{Meeson}},
  \bibinfo{author}{\bibfnamefont{M.}~\bibnamefont{Springford}},
  \bibnamefont{and} \bibinfo{author}{\bibfnamefont{G.}~\bibnamefont{Saito}},
  \bibinfo{journal}{Phys. Rev. B} \textbf{\bibinfo{volume}{65}},
  \bibinfo{pages}{064515} (\bibinfo{year}{2002}).

\bibitem[{\citenamefont{Heinecke and Winzer}(1995)}]{HeineckeW95}
\bibinfo{author}{\bibfnamefont{M.}~\bibnamefont{Heinecke}} \bibnamefont{and}
  \bibinfo{author}{\bibfnamefont{K.}~\bibnamefont{Winzer}},
  \bibinfo{journal}{Z. Phys. B-Condens. Mat.} \textbf{\bibinfo{volume}{98}},
  \bibinfo{pages}{147} (\bibinfo{year}{1995}).

\bibitem[{\citenamefont{Settai et~al.}(2001)\citenamefont{Settai, Shishido,
  Ikeda, Murakawa, Nakashima, Aoki, Haga, Harima, and
  Onuki}}]{SettaiSIMNAHHO01}
\bibinfo{author}{\bibfnamefont{R.}~\bibnamefont{Settai}},
  \bibinfo{author}{\bibfnamefont{H.}~\bibnamefont{Shishido}},
  \bibinfo{author}{\bibfnamefont{S.}~\bibnamefont{Ikeda}},
  \bibinfo{author}{\bibfnamefont{Y.}~\bibnamefont{Murakawa}},
  \bibinfo{author}{\bibfnamefont{M.}~\bibnamefont{Nakashima}},
  \bibinfo{author}{\bibfnamefont{D.}~\bibnamefont{Aoki}},
  \bibinfo{author}{\bibfnamefont{Y.}~\bibnamefont{Haga}},
  \bibinfo{author}{\bibfnamefont{H.}~\bibnamefont{Harima}}, \bibnamefont{and}
  \bibinfo{author}{\bibfnamefont{Y.}~\bibnamefont{Onuki}}, \bibinfo{journal}{J.
  Phys.-Condes. Matter} \textbf{\bibinfo{volume}{13}}, \bibinfo{pages}{L627}
  (\bibinfo{year}{2001}).

\bibitem[{\citenamefont{Inada et~al.}(1999)\citenamefont{Inada, Yamagami, Haga,
  Sakurai, Tokiwa, Honma, Yamamoto, Onuki, and Yanagisawa}}]{InadaYHSTHYOY99}
\bibinfo{author}{\bibfnamefont{Y.}~\bibnamefont{Inada}},
  \bibinfo{author}{\bibfnamefont{H.}~\bibnamefont{Yamagami}},
  \bibinfo{author}{\bibfnamefont{Y.}~\bibnamefont{Haga}},
  \bibinfo{author}{\bibfnamefont{K.}~\bibnamefont{Sakurai}},
  \bibinfo{author}{\bibfnamefont{Y.}~\bibnamefont{Tokiwa}},
  \bibinfo{author}{\bibfnamefont{T.}~\bibnamefont{Honma}},
  \bibinfo{author}{\bibfnamefont{E.}~\bibnamefont{Yamamoto}},
  \bibinfo{author}{\bibfnamefont{Y.}~\bibnamefont{Onuki}}, \bibnamefont{and}
  \bibinfo{author}{\bibfnamefont{T.}~\bibnamefont{Yanagisawa}},
  \bibinfo{journal}{J. Phys. Soc. Jpn.} \textbf{\bibinfo{volume}{68}},
  \bibinfo{pages}{3643} (\bibinfo{year}{1999}).

\bibitem[{\citenamefont{Ohkuni et~al.}(1999)\citenamefont{Ohkuni, Inada,
  Tokiwa, Sakurai, Settai, Honma, Haga, Yamamoto, Onuki, Yamagami, Takahashi,
  and Yanagisawa}}]{OhkuniITSSHHYOYTY99}
\bibinfo{author}{\bibfnamefont{H.}~\bibnamefont{Ohkuni}},
  \bibinfo{author}{\bibfnamefont{Y.}~\bibnamefont{Inada}},
  \bibinfo{author}{\bibfnamefont{Y.}~\bibnamefont{Tokiwa}},
  \bibinfo{author}{\bibfnamefont{K.}~\bibnamefont{Sakurai}},
  \bibinfo{author}{\bibfnamefont{R.}~\bibnamefont{Settai}},
  \bibinfo{author}{\bibfnamefont{T.}~\bibnamefont{Honma}},
  \bibinfo{author}{\bibfnamefont{Y.}~\bibnamefont{Haga}},
  \bibinfo{author}{\bibfnamefont{E.}~\bibnamefont{Yamamoto}},
  \bibinfo{author}{\bibfnamefont{Y.}~\bibnamefont{Onuki}},
  \bibinfo{author}{\bibfnamefont{H.}~\bibnamefont{Yamagami}},
  \bibinfo{author}{\bibfnamefont{S.}~\bibnamefont{Takahashi}},
  \bibnamefont{and}
  \bibinfo{author}{\bibfnamefont{T.}~\bibnamefont{Yanagisawa}},
  \bibinfo{journal}{Philos. Mag. B-Phys. Condens. Matter Stat. Mech. Electron.
  Opt.} \textbf{\bibinfo{volume}{79}}, \bibinfo{pages}{1045}
  (\bibinfo{year}{1999}).

\bibitem[{\citenamefont{Bergemann et~al.}(1997)\citenamefont{Bergemann, Julian,
  Mcmullan, Howard, Lonzarich, Lejay, Brison, and
  Flouquet}}]{BergemannJMHLLBF97}
\bibinfo{author}{\bibfnamefont{C.}~\bibnamefont{Bergemann}},
  \bibinfo{author}{\bibfnamefont{S.~R.} \bibnamefont{Julian}},
  \bibinfo{author}{\bibfnamefont{G.~J.} \bibnamefont{Mcmullan}},
  \bibinfo{author}{\bibfnamefont{B.~K.} \bibnamefont{Howard}},
  \bibinfo{author}{\bibfnamefont{G.~G.} \bibnamefont{Lonzarich}},
  \bibinfo{author}{\bibfnamefont{P.}~\bibnamefont{Lejay}},
  \bibinfo{author}{\bibfnamefont{J.~P.} \bibnamefont{Brison}},
  \bibnamefont{and} \bibinfo{author}{\bibfnamefont{J.}~\bibnamefont{Flouquet}},
  \bibinfo{journal}{Physica B} \textbf{\bibinfo{volume}{230}},
  \bibinfo{pages}{348} (\bibinfo{year}{1997}).

\bibitem[{\citenamefont{Hedo et~al.}(1995)\citenamefont{Hedo, Inada, Ishida,
  Yamamoto, Haga, Onuki, Higuchi, and Hasegawa}}]{HedoIIYHOHH95}
\bibinfo{author}{\bibfnamefont{M.}~\bibnamefont{Hedo}},
  \bibinfo{author}{\bibfnamefont{Y.}~\bibnamefont{Inada}},
  \bibinfo{author}{\bibfnamefont{T.}~\bibnamefont{Ishida}},
  \bibinfo{author}{\bibfnamefont{E.}~\bibnamefont{Yamamoto}},
  \bibinfo{author}{\bibfnamefont{Y.}~\bibnamefont{Haga}},
  \bibinfo{author}{\bibfnamefont{Y.}~\bibnamefont{Onuki}},
  \bibinfo{author}{\bibfnamefont{M.}~\bibnamefont{Higuchi}}, \bibnamefont{and}
  \bibinfo{author}{\bibfnamefont{A.}~\bibnamefont{Hasegawa}},
  \bibinfo{journal}{J. Phys. Soc. Jpn.} \textbf{\bibinfo{volume}{64}},
  \bibinfo{pages}{4535} (\bibinfo{year}{1995}).

\bibitem[{\citenamefont{Shoenberg}(1984)}]{shoenberg}
\bibinfo{author}{\bibfnamefont{D.}~\bibnamefont{Shoenberg}},
  \emph{\bibinfo{title}{Magnetic Oscillations in Metals}}
  (\bibinfo{publisher}{Cambridge University Press},
  \bibinfo{address}{Cambridge}, \bibinfo{year}{1984}).

\bibitem[{\citenamefont{Wasserman and Springford}(1994)}]{WassermanS94}
\bibinfo{author}{\bibfnamefont{A.}~\bibnamefont{Wasserman}} \bibnamefont{and}
  \bibinfo{author}{\bibfnamefont{M.}~\bibnamefont{Springford}},
  \bibinfo{journal}{Physica B} \textbf{\bibinfo{volume}{194}},
  \bibinfo{pages}{1801} (\bibinfo{year}{1994}).

\bibitem[{\citenamefont{Harima}(2002)}]{Harima02}
\bibinfo{author}{\bibfnamefont{H.}~\bibnamefont{Harima}},
  \bibinfo{journal}{Physica C} \textbf{\bibinfo{volume}{378}},
  \bibinfo{pages}{18} (\bibinfo{year}{2002}).

\bibitem[{\citenamefont{Rosner et~al.}(2002)\citenamefont{Rosner, An, Pickett,
  and Drechsler}}]{RosnerAPD02}
\bibinfo{author}{\bibfnamefont{H.}~\bibnamefont{Rosner}},
  \bibinfo{author}{\bibfnamefont{J.~M.} \bibnamefont{An}},
  \bibinfo{author}{\bibfnamefont{W.~E.} \bibnamefont{Pickett}},
  \bibnamefont{and} \bibinfo{author}{\bibfnamefont{S.~L.}
  \bibnamefont{Drechsler}}, \bibinfo{journal}{Phys. Rev. B}
  \textbf{\bibinfo{volume}{66}}, \bibinfo{pages}{024521}
  (\bibinfo{year}{2002}).

\bibitem[{\citenamefont{Mazin and Kortus}(2002)}]{MazinK02}
\bibinfo{author}{\bibfnamefont{I.~I.} \bibnamefont{Mazin}} \bibnamefont{and}
  \bibinfo{author}{\bibfnamefont{J.}~\bibnamefont{Kortus}},
  \bibinfo{journal}{Phys. Rev. B} \textbf{\bibinfo{volume}{65}},
  \bibinfo{pages}{180510} (\bibinfo{year}{2002}).

\bibitem[{\citenamefont{Yelland et~al.}(2002)\citenamefont{Yelland, Cooper,
  Carrington, Hussey, Meeson, Lee, Yamamoto, and Tajima}}]{YellandCCHMLYT02}
\bibinfo{author}{\bibfnamefont{E.~A.} \bibnamefont{Yelland}},
  \bibinfo{author}{\bibfnamefont{J.~R.} \bibnamefont{Cooper}},
  \bibinfo{author}{\bibfnamefont{A.}~\bibnamefont{Carrington}},
  \bibinfo{author}{\bibfnamefont{N.~E.} \bibnamefont{Hussey}},
  \bibinfo{author}{\bibfnamefont{P.~J.} \bibnamefont{Meeson}},
  \bibinfo{author}{\bibfnamefont{S.}~\bibnamefont{Lee}},
  \bibinfo{author}{\bibfnamefont{A.}~\bibnamefont{Yamamoto}}, \bibnamefont{and}
  \bibinfo{author}{\bibfnamefont{S.}~\bibnamefont{Tajima}},
  \bibinfo{journal}{Phys. Rev. Lett.} \textbf{\bibinfo{volume}{88}},
  \bibinfo{pages}{217002} (\bibinfo{year}{2002}).

\bibitem[{\citenamefont{Carrington et~al.}(2003)\citenamefont{Carrington,
  Meeson, Cooper, Balicas, Hussey, Yelland, Lee, Yamamoto, Tajima, Kazakov, and
  Karpinski}}]{CarringtonMCBHYLYTKK03}
\bibinfo{author}{\bibfnamefont{A.}~\bibnamefont{Carrington}},
  \bibinfo{author}{\bibfnamefont{P.~J.} \bibnamefont{Meeson}},
  \bibinfo{author}{\bibfnamefont{J.~R.} \bibnamefont{Cooper}},
  \bibinfo{author}{\bibfnamefont{L.}~\bibnamefont{Balicas}},
  \bibinfo{author}{\bibfnamefont{N.~E.} \bibnamefont{Hussey}},
  \bibinfo{author}{\bibfnamefont{E.~A.} \bibnamefont{Yelland}},
  \bibinfo{author}{\bibfnamefont{S.}~\bibnamefont{Lee}},
  \bibinfo{author}{\bibfnamefont{A.}~\bibnamefont{Yamamoto}},
  \bibinfo{author}{\bibfnamefont{S.}~\bibnamefont{Tajima}},
  \bibinfo{author}{\bibfnamefont{S.~M.} \bibnamefont{Kazakov}},
  \bibnamefont{and}
  \bibinfo{author}{\bibfnamefont{J.}~\bibnamefont{Karpinski}},
  \bibinfo{journal}{Phys. Rev. Lett.} \textbf{\bibinfo{volume}{91}},
  \bibinfo{pages}{037003} (\bibinfo{year}{2003}).

\bibitem[{pie()}]{piezolevers}
\bibinfo{note}{ThermoMicroscopes, Sunnyvale, California.
  \url{http:\\www.thermomicro.com}.}

\bibitem[{\citenamefont{Cooper et~al.}(2003)\citenamefont{Cooper, Carrington,
  Meeson, Yelland, Hussey, Balicas, Tajima, Lee, Kazakov, and
  Karpinski}}]{CooperCMYHBTLKK03}
\bibinfo{author}{\bibfnamefont{J.~R.} \bibnamefont{Cooper}},
  \bibinfo{author}{\bibfnamefont{A.}~\bibnamefont{Carrington}},
  \bibinfo{author}{\bibfnamefont{P.~J.} \bibnamefont{Meeson}},
  \bibinfo{author}{\bibfnamefont{E.~A.} \bibnamefont{Yelland}},
  \bibinfo{author}{\bibfnamefont{N.~E.} \bibnamefont{Hussey}},
  \bibinfo{author}{\bibfnamefont{L.}~\bibnamefont{Balicas}},
  \bibinfo{author}{\bibfnamefont{S.}~\bibnamefont{Tajima}},
  \bibinfo{author}{\bibfnamefont{S.}~\bibnamefont{Lee}},
  \bibinfo{author}{\bibfnamefont{S.~M.} \bibnamefont{Kazakov}},
  \bibnamefont{and}
  \bibinfo{author}{\bibfnamefont{J.}~\bibnamefont{Karpinski}},
  \bibinfo{journal}{Physica C} \textbf{\bibinfo{volume}{385}},
  \bibinfo{pages}{75} (\bibinfo{year}{2003}).

\bibitem[{\citenamefont{Karpinski et~al.}(2003)\citenamefont{Karpinski, Angst,
  Jun, Kazakov, Puzniak, Wisniewski, Roos, Keller, Perucchi, Degiorgi,
  Eskildsen, Bordet, Vinnikov, and Mironov}}]{KarpinskiAJKPWRKPDEBVM03}
\bibinfo{author}{\bibfnamefont{J.}~\bibnamefont{Karpinski}},
  \bibinfo{author}{\bibfnamefont{M.}~\bibnamefont{Angst}},
  \bibinfo{author}{\bibfnamefont{J.}~\bibnamefont{Jun}},
  \bibinfo{author}{\bibfnamefont{S.~M.} \bibnamefont{Kazakov}},
  \bibinfo{author}{\bibfnamefont{R.}~\bibnamefont{Puzniak}},
  \bibinfo{author}{\bibfnamefont{A.}~\bibnamefont{Wisniewski}},
  \bibinfo{author}{\bibfnamefont{J.}~\bibnamefont{Roos}},
  \bibinfo{author}{\bibfnamefont{H.}~\bibnamefont{Keller}},
  \bibinfo{author}{\bibfnamefont{A.}~\bibnamefont{Perucchi}},
  \bibinfo{author}{\bibfnamefont{L.}~\bibnamefont{Degiorgi}},
  \bibinfo{author}{\bibfnamefont{M.~R.} \bibnamefont{Eskildsen}},
  \bibinfo{author}{\bibfnamefont{P.}~\bibnamefont{Bordet}},
  \bibinfo{author}{\bibfnamefont{L.}~\bibnamefont{Vinnikov}}, \bibnamefont{and}
  \bibinfo{author}{\bibfnamefont{A.}~\bibnamefont{Mironov}},
  \bibinfo{journal}{Supercond. Sci. Technol.} \textbf{\bibinfo{volume}{16}},
  \bibinfo{pages}{221} (\bibinfo{year}{2003}).

\bibitem[{\citenamefont{Hao and Clem}(1991)}]{HaoC91}
\bibinfo{author}{\bibfnamefont{Z.~D.} \bibnamefont{Hao}} \bibnamefont{and}
  \bibinfo{author}{\bibfnamefont{J.~R.} \bibnamefont{Clem}},
  \bibinfo{journal}{Phys. Rev. B} \textbf{\bibinfo{volume}{43}},
  \bibinfo{pages}{7622} (\bibinfo{year}{1991}).

\bibitem[{\citenamefont{Angst et~al.}(2002)\citenamefont{Angst, Puzniak,
  Wisniewski, Jun, Kazakov, Karpinski, Roos, and Keller}}]{AngstPWJKKRK02}
\bibinfo{author}{\bibfnamefont{M.}~\bibnamefont{Angst}},
  \bibinfo{author}{\bibfnamefont{R.}~\bibnamefont{Puzniak}},
  \bibinfo{author}{\bibfnamefont{A.}~\bibnamefont{Wisniewski}},
  \bibinfo{author}{\bibfnamefont{J.}~\bibnamefont{Jun}},
  \bibinfo{author}{\bibfnamefont{S.~M.} \bibnamefont{Kazakov}},
  \bibinfo{author}{\bibfnamefont{J.}~\bibnamefont{Karpinski}},
  \bibinfo{author}{\bibfnamefont{J.}~\bibnamefont{Roos}}, \bibnamefont{and}
  \bibinfo{author}{\bibfnamefont{H.}~\bibnamefont{Keller}},
  \bibinfo{journal}{Phys. Rev. Lett.} \textbf{\bibinfo{volume}{88}},
  \bibinfo{pages}{167004} (\bibinfo{year}{2002}).

\bibitem[{\citenamefont{Angst et~al.}(2003)\citenamefont{Angst, Puzniak,
  Wisniewski, Roos, Keller, Miranovic, Jun, Kazakov, and
  Karpinski}}]{AngstPWRKMJKK03}
\bibinfo{author}{\bibfnamefont{M.}~\bibnamefont{Angst}},
  \bibinfo{author}{\bibfnamefont{R.}~\bibnamefont{Puzniak}},
  \bibinfo{author}{\bibfnamefont{A.}~\bibnamefont{Wisniewski}},
  \bibinfo{author}{\bibfnamefont{J.}~\bibnamefont{Roos}},
  \bibinfo{author}{\bibfnamefont{H.}~\bibnamefont{Keller}},
  \bibinfo{author}{\bibfnamefont{P.}~\bibnamefont{Miranovic}},
  \bibinfo{author}{\bibfnamefont{J.}~\bibnamefont{Jun}},
  \bibinfo{author}{\bibfnamefont{S.~M.} \bibnamefont{Kazakov}},
  \bibnamefont{and}
  \bibinfo{author}{\bibfnamefont{J.}~\bibnamefont{Karpinski}},
  \bibinfo{journal}{Physica C} \textbf{\bibinfo{volume}{385}},
  \bibinfo{pages}{143} (\bibinfo{year}{2003}).

\bibitem[{\citenamefont{Rydh et~al.}(2003)\citenamefont{Rydh, Welp, Hiller,
  Koshelev, Kwok, Crabtree, Kim, Kim, Jung, Lee, Kang, and
  Lee}}]{RydhWHKKCKKJLKL03}
\bibinfo{author}{\bibfnamefont{A.}~\bibnamefont{Rydh}},
  \bibinfo{author}{\bibfnamefont{U.}~\bibnamefont{Welp}},
  \bibinfo{author}{\bibfnamefont{J.~M.} \bibnamefont{Hiller}},
  \bibinfo{author}{\bibfnamefont{A.~E.} \bibnamefont{Koshelev}},
  \bibinfo{author}{\bibfnamefont{W.~K.} \bibnamefont{Kwok}},
  \bibinfo{author}{\bibfnamefont{G.~W.} \bibnamefont{Crabtree}},
  \bibinfo{author}{\bibfnamefont{K.~H.~P.} \bibnamefont{Kim}},
  \bibinfo{author}{\bibfnamefont{K.~H.} \bibnamefont{Kim}},
  \bibinfo{author}{\bibfnamefont{C.~U.} \bibnamefont{Jung}},
  \bibinfo{author}{\bibfnamefont{H.~S.} \bibnamefont{Lee}},
  \bibinfo{author}{\bibfnamefont{B.}~\bibnamefont{Kang}}, \bibnamefont{and}
  \bibinfo{author}{\bibfnamefont{S.~I.} \bibnamefont{Lee}},
  \bibinfo{journal}{Phys. Rev. B} \textbf{\bibinfo{volume}{68}},
  \bibinfo{pages}{172502} (\bibinfo{year}{2003}).

\bibitem[{\citenamefont{Terashima et~al.}(1997)\citenamefont{Terashima,
  Haworth, Takeya, Uji, Aoki, and Kadowaki}}]{TerashimaHTUAK97}
\bibinfo{author}{\bibfnamefont{T.}~\bibnamefont{Terashima}},
  \bibinfo{author}{\bibfnamefont{C.}~\bibnamefont{Haworth}},
  \bibinfo{author}{\bibfnamefont{H.}~\bibnamefont{Takeya}},
  \bibinfo{author}{\bibfnamefont{S.}~\bibnamefont{Uji}},
  \bibinfo{author}{\bibfnamefont{H.}~\bibnamefont{Aoki}}, \bibnamefont{and}
  \bibinfo{author}{\bibfnamefont{K.}~\bibnamefont{Kadowaki}},
  \bibinfo{journal}{Phys. Rev. B} \textbf{\bibinfo{volume}{56}},
  \bibinfo{pages}{5120} (\bibinfo{year}{1997}).

\bibitem[{\citenamefont{Janssen et~al.}(1998)\citenamefont{Janssen, Haworth,
  Hayden, Meeson, Springford, and Wasserman}}]{JanssenHHMSW98}
\bibinfo{author}{\bibfnamefont{T.~J. B.~M.} \bibnamefont{Janssen}},
  \bibinfo{author}{\bibfnamefont{C.}~\bibnamefont{Haworth}},
  \bibinfo{author}{\bibfnamefont{S.~M.} \bibnamefont{Hayden}},
  \bibinfo{author}{\bibfnamefont{P.}~\bibnamefont{Meeson}},
  \bibinfo{author}{\bibfnamefont{M.}~\bibnamefont{Springford}},
  \bibnamefont{and}
  \bibinfo{author}{\bibfnamefont{A.}~\bibnamefont{Wasserman}},
  \bibinfo{journal}{Phys. Rev. B} \textbf{\bibinfo{volume}{57}},
  \bibinfo{pages}{11698} (\bibinfo{year}{1998}).

\bibitem[{\citenamefont{Goll et~al.}(1996)\citenamefont{Goll, Heinecke, Jansen,
  Joss, Nguyen, Steep, Winzer, and Wyder}}]{GollHJJNSWW96}
\bibinfo{author}{\bibfnamefont{G.}~\bibnamefont{Goll}},
  \bibinfo{author}{\bibfnamefont{M.}~\bibnamefont{Heinecke}},
  \bibinfo{author}{\bibfnamefont{A.~G.~M.} \bibnamefont{Jansen}},
  \bibinfo{author}{\bibfnamefont{W.}~\bibnamefont{Joss}},
  \bibinfo{author}{\bibfnamefont{L.}~\bibnamefont{Nguyen}},
  \bibinfo{author}{\bibfnamefont{E.}~\bibnamefont{Steep}},
  \bibinfo{author}{\bibfnamefont{K.}~\bibnamefont{Winzer}}, \bibnamefont{and}
  \bibinfo{author}{\bibfnamefont{P.}~\bibnamefont{Wyder}},
  \bibinfo{journal}{Phys. Rev. B} \textbf{\bibinfo{volume}{53}},
  \bibinfo{pages}{R8871} (\bibinfo{year}{1996}).

\bibitem[{\citenamefont{Carrington and Fletcher}(2004)}]{carringtonF04}
\bibinfo{author}{\bibfnamefont{A.}~\bibnamefont{Carrington}} \bibnamefont{and}
  \bibinfo{author}{\bibfnamefont{J.}~\bibnamefont{Fletcher}}
  (\bibinfo{year}{2004}).

\bibitem[{\citenamefont{Boaknin et~al.}(2003)\citenamefont{Boaknin, Tanatar,
  Paglione, Hawthorn, Ronning, Hill, Sutherland, Taillefer, Sonier, Hayden, and
  Brill}}]{BoakninTPHRHSTSHB03}
\bibinfo{author}{\bibfnamefont{E.}~\bibnamefont{Boaknin}},
  \bibinfo{author}{\bibfnamefont{M.~A.} \bibnamefont{Tanatar}},
  \bibinfo{author}{\bibfnamefont{J.}~\bibnamefont{Paglione}},
  \bibinfo{author}{\bibfnamefont{D.}~\bibnamefont{Hawthorn}},
  \bibinfo{author}{\bibfnamefont{F.}~\bibnamefont{Ronning}},
  \bibinfo{author}{\bibfnamefont{R.~W.} \bibnamefont{Hill}},
  \bibinfo{author}{\bibfnamefont{M.}~\bibnamefont{Sutherland}},
  \bibinfo{author}{\bibfnamefont{L.}~\bibnamefont{Taillefer}},
  \bibinfo{author}{\bibfnamefont{J.}~\bibnamefont{Sonier}},
  \bibinfo{author}{\bibfnamefont{S.~M.} \bibnamefont{Hayden}},
  \bibnamefont{and} \bibinfo{author}{\bibfnamefont{J.~W.} \bibnamefont{Brill}},
  \bibinfo{journal}{Phys. Rev. Lett.} \textbf{\bibinfo{volume}{90}},
  \bibinfo{pages}{117003} (\bibinfo{year}{2003}).

\bibitem[{\citenamefont{V.~Guritanu et~al.}(2004)\citenamefont{V.~Guritanu,
  Goldacker, Bouquet, Wang, Lortz, Goll, and Junod}}]{GuritanuGBWLGJ04}
\bibinfo{author}{\bibfnamefont{V.}~\bibnamefont{V.~Guritanu}},
  \bibinfo{author}{\bibfnamefont{W.}~\bibnamefont{Goldacker}},
  \bibinfo{author}{\bibfnamefont{F.}~\bibnamefont{Bouquet}},
  \bibinfo{author}{\bibfnamefont{Y.}~\bibnamefont{Wang}},
  \bibinfo{author}{\bibfnamefont{R.}~\bibnamefont{Lortz}},
  \bibinfo{author}{\bibfnamefont{G.}~\bibnamefont{Goll}}, \bibnamefont{and}
  \bibinfo{author}{\bibfnamefont{A.}~\bibnamefont{Junod}}
  (\bibinfo{year}{2004}), \bibinfo{note}{cond-mat/0403590}.

\bibitem[{\citenamefont{Yasui and Kita}(2002)}]{YasuiK02}
\bibinfo{author}{\bibfnamefont{K.}~\bibnamefont{Yasui}} \bibnamefont{and}
  \bibinfo{author}{\bibfnamefont{T.}~\bibnamefont{Kita}},
  \bibinfo{journal}{Phys. Rev. B} \textbf{\bibinfo{volume}{66}},
  \bibinfo{pages}{184516} (\bibinfo{year}{2002}).

\bibitem[{\citenamefont{Maki}(1991)}]{Maki91}
\bibinfo{author}{\bibfnamefont{K.}~\bibnamefont{Maki}}, \bibinfo{journal}{Phys.
  Rev. B} \textbf{\bibinfo{volume}{44}}, \bibinfo{pages}{2861}
  (\bibinfo{year}{1991}).

\bibitem[{\citenamefont{Brandt et~al.}(1967)\citenamefont{Brandt, Pesch, and
  Tewordt}}]{BrandtPT67}
\bibinfo{author}{\bibfnamefont{U.}~\bibnamefont{Brandt}},
  \bibinfo{author}{\bibfnamefont{W.}~\bibnamefont{Pesch}}, \bibnamefont{and}
  \bibinfo{author}{\bibfnamefont{L.}~\bibnamefont{Tewordt}},
  \bibinfo{journal}{Z. Phys} \textbf{\bibinfo{volume}{201}},
  \bibinfo{pages}{209} (\bibinfo{year}{1967}).

\bibitem[{\citenamefont{Stephen}(1992)}]{Stephen92}
\bibinfo{author}{\bibfnamefont{M.~J.} \bibnamefont{Stephen}},
  \bibinfo{journal}{Phys. Rev. B} \textbf{\bibinfo{volume}{45}},
  \bibinfo{pages}{5481} (\bibinfo{year}{1992}).

\bibitem[{\citenamefont{Miyake}(1993)}]{Miyake93}
\bibinfo{author}{\bibfnamefont{K.}~\bibnamefont{Miyake}},
  \bibinfo{journal}{Physica B} \textbf{\bibinfo{volume}{188}},
  \bibinfo{pages}{115} (\bibinfo{year}{1993}).

\bibitem[{\citenamefont{Miller and Gyorffy}(1995)}]{MillerG95}
\bibinfo{author}{\bibfnamefont{P.}~\bibnamefont{Miller}} \bibnamefont{and}
  \bibinfo{author}{\bibfnamefont{B.~L.} \bibnamefont{Gyorffy}},
  \bibinfo{journal}{J. Phys.-Condes. Matter} \textbf{\bibinfo{volume}{7}},
  \bibinfo{pages}{5579} (\bibinfo{year}{1995}).

\bibitem[{mas()}]{massnote}
\bibinfo{note}{This scaling is found both experimentally and in band-structure
  calculations \cite{Harima02}.}

\bibitem[{\citenamefont{Mazin et~al.}(2004)\citenamefont{Mazin, Andersen,
  Jepsen, Golubov, Dolgov, and Kortus}}]{MazinAJGDK04}
\bibinfo{author}{\bibfnamefont{I.~I.} \bibnamefont{Mazin}},
  \bibinfo{author}{\bibfnamefont{O.~K.} \bibnamefont{Andersen}},
  \bibinfo{author}{\bibfnamefont{O.}~\bibnamefont{Jepsen}},
  \bibinfo{author}{\bibfnamefont{A.~A.} \bibnamefont{Golubov}},
  \bibinfo{author}{\bibfnamefont{O.~V.} \bibnamefont{Dolgov}},
  \bibnamefont{and} \bibinfo{author}{\bibfnamefont{J.}~\bibnamefont{Kortus}},
  \bibinfo{journal}{Phys. Rev. B} \textbf{\bibinfo{volume}{69}},
  \bibinfo{pages}{056501} (\bibinfo{year}{2004}).

\bibitem[{\citenamefont{Duncan and Gyorffy}(2003)}]{DuncanG03}
\bibinfo{author}{\bibfnamefont{K.~P.} \bibnamefont{Duncan}} \bibnamefont{and}
  \bibinfo{author}{\bibfnamefont{B.~L.} \bibnamefont{Gyorffy}},
  \bibinfo{journal}{J. Phys.-Condes. Matter} \textbf{\bibinfo{volume}{15}},
  \bibinfo{pages}{239} (\bibinfo{year}{2003}).

\end{thebibliography}
\end{document}